\documentclass[floatfix,preprint,a4paper,aps,prd,preprintnumbers,amsfonts,amsmath]{revtex4}
\usepackage{geometry}
\DeclareMathAlphabet{\scr}{U}{rsfs}{m}{n}
\usepackage{epsfig}
\usepackage[dvips]{color}
\input{axodraw.sty}

\usepackage{graphicx, graphics}
\usepackage{latexsym}
\usepackage{mathrsfs,amssymb}
\usepackage{psfrag}
\usepackage{hepparticles}
\usepackage{hepnicenames}
\usepackage{heppennames}
\usepackage{braket}

\usepackage{fullpage}
\usepackage{amsmath}
\usepackage{dcolumn}
\usepackage{dsfont}

\newcommand{\columnspacefixer}[0]{\hspace{-4mm}\!\!\!\!}

\newcommand{\newhline}[0]{\\[1mm] \hline}
\newcommand{\slashed}[1]{\ensuremath{{#1}{\!}{\!}{\!}{\!}{\:}/}}
\newcommand{\Slashed}[1]{\ensuremath{{#1}{\!}{\!}{\!}{\!}{\!}{\:}/}}
\newcommand{\gammabar}[0]{\ensuremath{{\gamma}^{5}}}
\newcommand{\dotproduct}[2]{\ensuremath{{#1}{\!}{\!}{\:}{\!}{\:}{\cdot}{\!}{\!}{\:}{\!}{\:}{\!}{\:}{#2}}}
\newcommand{\superpartner}[1]{\ensuremath{{\tilde{{#1}}}}}
\newcommand{\xone}[0]{\ensuremath{{{\superpartner{{\chi}}}_{1}^{0}}}}
\newcommand{\xoneL}[0]{\ensuremath{{{\superpartner{{\chi}}}_{1L}^{0}}}}
\newcommand{\xbarone}[0]{\ensuremath{{{\bar{{\superpartner{{\chi}}}}}_{1}^{0}}}}
\newcommand{\xbaroneL}[0]{\ensuremath{{{\bar{{\superpartner{{\chi}}}}}_{1L}^{0}}}}
\newcommand{\mxone}[0]{\ensuremath{m_{{{\tilde{\chi}}^{0}_{1}}}}}
\newcommand{\mxonesq}[0]{\ensuremath{m_{{{\tilde{\chi}}^{0}_{1}}}^{2}}}
\newcommand{\neuino}[1]{\ensuremath{{{\superpartner{{\chi}}}_{#1}^{0}}}}

\newcommand{\FormCalc}[0]{\texttt{FormCalc}}

\newcommand{\SOFTSUSY}[0]{\texttt{SOFTSUSY}}
\newcommand{\MadGraph}[0]{\texttt{MadGraph}}
\newcommand{\Ktopi}[0]{\ensuremath{{K^{-} {\to} {\pi}^{-} {\xone} {\xone}}}}
\newcommand{\BtoK}[0]{\ensuremath{{B^{-} {\to} K^{-} {\xone} {\xone}}}}
\newcommand{\eg}[0]{\textit{e.g.}}
\newcommand{\ie}[0]{\textit{i.e.}}

\newcommand{\myatop}[2]{\ensuremath{{{#1}\atop{#2}}}}

\def\lsim{\raise0.3ex\hbox{$\;<$\kern-0.75em\raise-1.1ex\hbox{$\sim\;$}}}
\def\gsim{\raise0.3ex\hbox{$\;>$\kern-0.75em\raise-1.1ex\hbox{$\sim\;$}}}

\begin{document}

\preprint{Bonn-TH-2009-04}
\preprint{DESY 09-068}
\preprint{PITHA 09/10}
\preprint{QMUL-PH-09-11}

\vspace*{10mm}

\title{Rare meson decays into very light neutralinos\vspace*{-5mm}}

\author{H.~K.~Dreiner} \email[]{dreiner@th.physik.uni-bonn.de}
\affiliation{Bethe Center for Theoretical Physics \& Physikalisches
  Institut der Universit{\"{a}}t Bonn, Nu{\ss}allee 12, 53115 Bonn,
  Germany}

\author{S.~Grab} \email[]{sgrab@th.physik.uni-bonn.de}
\affiliation{Bethe Center for Theoretical Physics \& Physikalisches
  Institut der Universit{\"{a}}t Bonn, Nu{\ss}allee 12, 53115 Bonn,
  Germany}

\author{Daniel Koschade} \email[]{d.koschade@qmul.ac.uk}
\affiliation{Institut f{\"{u}}r Theoretische Physik, RWTH Aachen
  University, 52056 Aachen, Germany} \affiliation{\mbox{Centre for Research
  in String Theory, Department of Physics}, Queen Mary, University of
  London, E1 4NS London, United Kingdom}

\author{M.~Kr{\"{a}}mer} \email[]{mkraemer@physik.rwth-aachen.de}
\affiliation{Institut f{\"{u}}r Theoretische Physik, RWTH Aachen
  University, 52056 Aachen, Germany}

\author{Ulrich Langenfeld} \email[]{ulrichl@ifh.de}
\affiliation{DESY, Platanenallee 6, D-15738 Zeuthen, Germany}

\author{Ben O'Leary} \email[]{oleary@physik.rwth-aachen.de}
\affiliation{Institut f{\"{u}}r Theoretische Physik, RWTH Aachen
  University, 52056 Aachen, Germany}

\begin{abstract}
  We investigate the bounds on the mass of the lightest neutralino
  from rare meson decays within the MSSM with and without minimal
  flavor violation.  We present explicit formulae for the two-body
  decays of mesons into light neutralinos and perform the first
  complete calculation of the loop-induced decays of kaons
  to pions and light neutralinos and $B$ mesons to kaons and light
  neutralinos. We find that the supersymmetric branching ratios are
  strongly suppressed within the MSSM with minimal flavor violation,
  and that no bounds on the neutralino mass can be inferred from
  experimental data, \ie~a massless neutralino is allowed.  The
  branching ratios for kaon and $B$ meson decays into light
  neutralinos may, however, be enhanced when one allows for
  non-minimal flavor violation. We find new constraints on the MSSM
  parameter space for such scenarios and discuss prospects for
  future kaon and $B$ meson experiments.  Finally, we comment on the
  search for light neutralinos in monojet signatures at the Tevatron
  and at the LHC.

\end{abstract}

\vspace*{-9mm}

\maketitle


\section{Introduction }

Supersymmetry (SUSY) is an attractive candidate for physics beyond the
standard model (SM)~\cite{susy1,susy2,susy3}. In the minimal
supersymmetric extension of the SM (MSSM) the superpartners of the
electroweak gauge and Higgs bosons (the gauginos and higgsinos) mix to
form two electrically charged and four electrically neutral mass
eigenstates, called charginos, $\tilde{\chi}^{\pm}_{1,2}$, and
neutralinos, $\tilde{\chi}^{0}_{1,2,3,4}$, respectively. In many
supersymmetric models, the neutralino $\tilde{\chi}^{0}_{1}$ is the
lightest supersymmetric particle (LSP) and thus plays a special role
in phenomenology. In particular, in models with conserved proton
hexality~\cite{Dreiner:2005rd} (or conserved
$R$-parity~\cite{Farrar:1978xj}) a neutralino LSP is stable and provides
a promising dark matter
candidate~\cite{Ellis:1983ew, Pagels:1981ke, Goldberg:1983nd}.

So far, no evidence for supersymmetric particles has been found, and
lower limits on sparticle masses have been derived from searches at
colliders and from precision measurements of low-energy observables.
The particle data group (PDG)~\cite{PDG} quotes a lower limit on the
mass of the lightest neutralino~\cite{Abdallah:2003xe}
\begin{equation}\label{eq:neu_mass_bound}
m_{\neuino{1} } > 46~{\rm GeV}\quad \mbox{(95\% C.L.)}.
\end{equation}
This limit, however, is not model independent but assumes the
unification of the gaugino mass parameters $M_1$ and $M_2$ at some
high energy scale. The renormalization group evolution of the mass
parameters then implies
\begin{equation}\label{eq:m_uni}
M_1 = \frac{5}{3}\frac{{g'}^2}{g^2}M_2 = \frac{5}{3}\tan^2\theta_W M_2
\approx \frac{1}{2}M_2
\end{equation}
at the electroweak scale, where $g$ and $g'$ denote the $SU(2)_L$ and
$U(1)_Y$ gauge couplings, respectively, and $\theta_W$ is the weak
mixing angle.  The experimental mass bound for the lighter chargino,
$m_{\tilde{\chi}^{\pm}_1} > 94~{\rm GeV}$ ($95\%$ C.L.)~\cite{PDG}
places a lower bound on $M_2$ (and on the higgsino mass parameter
$\mu$) and indirectly, through Eq.~(\ref{eq:m_uni}), on $M_1$. The
bounds on $M_1$, $M_2$ and $\mu$, as well as the lower bound on
$\tan\beta \equiv v_2/v_1$ (the ratio of the two vacuum
expectation values of the MSSM Higgs doublets) from the LEP Higgs
searches \cite{lhwg}, then in turn give rise to the lower bound on
the mass of the lightest neutralino, Eq.~(\ref{eq:neu_mass_bound})
above.

In this paper, we investigate a more general MSSM scenario where the
assumption Eq.~(\ref{eq:m_uni}) is dropped and the gaugino mass
parameters $M_1$ and $M_2$ are treated as independent parameters. It
has been shown that in such a scenario, one can adjust $M_1$ and $M_2$
such that the lightest neutralino is
massless~\cite{Bartl:1989ms, Gogoladze:2002xp,Dreiner:2003wh}.
Specifically, $m_{\neuino{1}} = 0$ if
\begin{equation}\label{eq:m1_min}
  M_1 =
\frac{M_2 m_Z^2 \sin(2\beta)\sin^2\theta_W}{\mu M_2
- m_Z^2 \sin(2\beta) \cos\theta_W}
  \approx \frac{m_Z^2 \sin(2\beta)\sin^2\theta_W}{\mu}
  \approx 2.5~{\rm GeV}\left(\frac{10}{\tan\beta}\right)
  \left(\frac{150~{\rm GeV}}{\mu}\right),
\end{equation}
where $m_Z$ is the $Z$ gauge boson mass.  The lower bound on the
lightest chargino mass, $m_{\tilde{\chi}^{\pm}_1} > 94~{\rm GeV}$,
leads to lower bounds of $|\mu|, M_2 \gsim
100$~GeV~\cite{Barger:2005hb}. Thus, adjusting $M_1$ according to
Eq.~(\ref{eq:m1_min}) so that $m_{\neuino{1} } = 0$ implies $M_1 \ll
M_2, \mu$. In this region of parameter space, the bino component of
the lightest neutralino is above 90\%, \ie\ the lightest neutralino
couples predominantly to
hypercharge~\cite{Dreiner:2003wh,Dreiner:2007fw}.

Note that choosing a renormalization scheme in which $m_{\neuino{1}}$
is an input parameter~\cite{Fritzsche:2002bi,Dreiner:2009ic} implies
that a small ${\xone}$ mass at tree level will remain small also after
radiative corrections.  However, note that a small or zero neutralino
mass can not always be obtained in the presence of non-zero complex
phases for $M_1$ and $\mu$~\cite{Dreiner:2009ic}.  (One can always
choose a convention where $M_2$ is real by absorbing its phase into a
redefinition of the gaugino fields.)

The phenomenology of a very light or massless neutralino has been
discussed by many previous authors, in particular with respect to its
astrophysical, see \textit{e.g.} 
Refs.~\cite{Ellis:1988aa,Choudhury:1999tn,Kachelriess:2000dz,Dreiner:2003wh},
and cosmological implications, see \textit{e.g.} 
\cite{Hooper:2002nq,Bottino:2003iu,Belanger:2003wb,Gunion:2005rw,Profumo:2008yg,Dreiner:2009ic}. Light
neutralinos have also been discussed in the context of collider searches
\cite{Fayet:1982ky,Ellis:1982zz,Grassie:1983kq,Barger:2005hb,Barger:2007nv,Dreiner:2007vm,Dreiner:2006sb},
as well as electroweak precision observables
\cite{Chankowski:1993eu,Dreiner:2009ic} and rare meson decays
\cite{Gaillard:1982rw,Ellis:1982ve,Kobayashi:1984wu,Nieves:1985ir,Dobroliubov:1987cba,Dreiner:2009ic}.
In a previous paper~\cite{Dreiner:2009ic}, some of the present authors
have presented an extensive study of these phenomenological
constraints.  While a very light or massless neutralino cannot provide
the cold dark matter content of the universe, it is consistent with
all existing laboratory constraints and astrophysical and cosmological
observations. For a comprehensive review of the literature on very
light neutralinos, we refer to Ref.~\cite{Dreiner:2009ic}.

In the present work we focus on a specific aspect of the phenomenology
of light neutralinos, namely rare meson decays to light neutralinos.
For more than three decades, rare meson decays have been considered a
sensitive test for the presence of new physics. In this paper, we
investigate the tree-level decay of pseudoscalar and vector mesons and
the loop-induced decay of kaons to pions and light neutralinos
and $B$ mesons to kaons and light neutralinos
within the MSSM. A review and an update of the literature has been
presented in a recent accompanying paper Ref.~\cite{Dreiner:2009ic}.
We go beyond Ref.~\cite{Dreiner:2009ic} and the previous work in
several aspects. First, we provide details of the calculations and
explicit formulae for the two-body decays of mesons $M$ into light
neutralinos, $M \to {\xone} {\xone}$, that will allow one to estimate
the decay width $\Gamma(M \to {\xone} {\xone})$ also in generic
supersymmetric theories beyond the MSSM.  Moreover, we perform the
first complete calculation of the loop-induced decays ${\Ktopi}$ and
${\BtoK}$ in the MSSM with minimal flavor violation
(MFV)~\cite{Gabrielli:1994ff}. We show that these decays are strongly
suppressed with respect to the SM decays $K^- \to \pi^- \nu \bar{\nu}$
and $B^- \to K^- \nu \bar{\nu}$. We also consider non-minimal flavor
violation in the MSSM and demonstrate that rare kaon and $B$ meson
decays do provide new constraints on the supersymmetric parameter
space in the case of light or massless neutralinos.  Finally, we
comment on the direct search for light neutralinos in monojet
signatures at the Tevatron and at the LHC.

We note that this MSSM with extremely light neutralinos is not the only model which allows for invisible meson decays into beyond-the-SM (BSM) particles.  Other models can have similar decays, which may be within reach of near-future experiments.  The standard model can be simply extended to accommodate light gauge singlet scalars which are dark matter candidates \cite{Silveira:1985rk,McDonald:1993ex,Burgess:2000yq}, though it requires some fine-tuning.  Rare meson decays to the scalars in this model have been calculated in Refs.~\cite{Bird:2004ts,Bird:2006jd}.  These decays in more complex, less fine-tuned models \cite{Boehm:2002yz,Boehm:2003hm}, in which the light scalars couple to new heavy fermions or a new light $U( 1 )^{{\prime}}$ gauge boson, have been calculated in Refs.~\cite{McElrath:2005bp,Fayet:2006sp,Fayet:2007ua}.  Models with extra gauged $U( 1 )$ symmetries also allow for light fermionic dark matter candidates \cite{Fayet:2004bw,Fayet:2004kz}, and the rare meson decays into these fermions have been calculated in Refs.~\cite{Fayet:2006sp,Fayet:2007ua}.  The next-to-minimal supersymmetric standard model (NMSSM) also allows for light neutralinos, and in the NMSSM they remain a dark matter candidate \cite{Gunion:2005rw}; the NMSSM also admits a light pseudoscalar dark matter candidate.  The rare meson decays into invisible BSM particles in the NMSSM have been calculated in Refs.~\cite{Hiller:2004ii,Bird:2006jd}.  Another supersymmetric model is that of Ref.~\cite{Feng:2008ya}, a model with gauge mediation of supersymmetry breaking with hidden sectors, and in these hidden sectors reside candidates for light dark matter.  Decays of mesons into these dark matter candidates have been calculated in Ref.~\cite{McKeen:2009rm}.

The paper is structured as follows. In Sect.~\ref{sectionII} we
present the calculation of the decay of pseudoscalar and vector mesons
to light neutralinos and compare the results with existing
experimental bounds.  Since the decay amplitudes involve virtual
squarks which, in the MSSM, are generically heavy, $m_{\tilde{q}}\gsim
100$~GeV, the branching ratios to neutralinos are small, well below
the experimental upper bounds. We provide explicit formulae which
allow one to study the dependence of the decay rate on the generic
supersymmetric masses and mixings. In Sect.~\ref{MFV_section} we
discuss the complete calculation of the loop-induced decays ${\Ktopi}$
and ${\BtoK}$ in the MSSM with minimal flavor violation. We present
numerical results for various supersymmetric scenarios and find that
the branching ratios are several orders of magnitude smaller than the
SM processes with neutrinos instead of neutralinos in the final state.
The branching ratios for the ${\Ktopi}$ and ${\BtoK}$ decays may,
however, be greatly enhanced when one allows for non-minimal flavor
violation in the MSSM, as discussed in Sect.~\ref{sec:non-mfv}. We
find new constraints on the MSSM parameter space and argue that future
experiments in the kaon and $B$ meson sector may be able to probe
light neutralinos with masses up to approximately $1$~GeV
through rare decays.  We comment on direct searches for light
neutralinos with a monojet signature at hadron colliders in
Sect.~\ref{sec:monojet} and conclude in Sect.~\ref{sec:conclusion}.
The appendix provides details on the form factors needed for the
calculation of the kaon and $B$ meson decays.

\section{Meson decays into light neutralinos}
\label{sectionII}

In this section, we consider the decays of the electrically
neutral, flavorless mesons
$M = {\pi}^{0}, {\eta}, {{\eta}'}, {\rho}^{0}, {\omega}, {\phi},
J/{\psi}, {\Upsilon}$ into a pair of light bino-like
neutralinos, $M {\to} {\xone} {\xone}$.  We are thus considering the
decay of an invisible particle into other invisible particles,
analogous to meson decays into neutrinos in the SM. Such decays have
been studied and searched for experimentally, making clever use of
kinematics in decay chains.  For example, an
invisibly-decaying ${\pi}^{0}$ can be reconstructed from a
$K^{+} {\to} {\pi}^{+} {\pi}^{0}$ decay \cite{Artamonov:2005cu},
while an invisibly-decaying $J/{\psi}$ may be observed by tagging on
the invariant mass of the di-pion system in the decay
${\psi}( 2S ) {\to} J/{\psi} + {\pi} {\pi}$~\cite{Rich:1976nu}.
The decays of pseudoscalar and vector mesons to light photino-like
neutralinos have first been studied in
Refs.~\cite{susy3, Gaillard:1982rw, Ellis:1982ve, Ellis:1983ed,
Campbell:1983jx, Dobroliubov:1987cba}.  
More recent calculations
addressing $B$ meson and ${\Upsilon}$ decays into light new matter in
the MSSM and beyond can be found in Refs.~\cite{Adhikari:1994iv,
Adhikari:1994wh, Chang:1997tq, McElrath:2005bp}.  We comment on the
comparison with existing results when discussing the individual decays
below.

Note that while decays with an additional photon in the final state
such as ${\Upsilon} {\to} {\xone} {\xone} + {\gamma}$ are
experimentally easier to detect, they suffer from a further
suppression by a factor of ${\alpha}$.  Since the branching ratios
we obtain for the two-body decays $M {\to} {\xone} {\xone}$ are
extremely small anyway, we do not pursue the radiative decays
$M {\to} {\xone} {\xone} + {\gamma}$.

\begin{figure}
\begin{center}
\includegraphics[width=8cm]{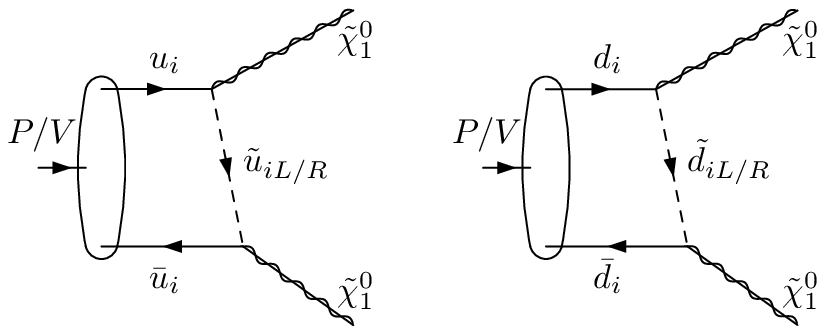}
\caption{Feynman diagrams for the decay $P/V {\to} {\xone} {\xone}$,
  omitting diagrams with the ${\xone}$s crossed.}
\label{PV_to_binos_diagram}
\end{center}
\end{figure}

The tree-level contributions to the decays of pseudoscalar $(P)$ or
vector $(V)$ mesons, $P/V {\to} {\xone} {\xone}$, are mediated by
${\superpartner{u}}_{{iL/R}}$ and ${\superpartner{d}}_{{iL/R}}$
$t$-channel exchange, as shown in Fig.~\ref{PV_to_binos_diagram}.
Here ${\superpartner{u}}_{{iL/R}}$/${\superpartner{d}}_{{iL/R}}$
denote a left- or right-handed up or down squark of flavor~$i$.

\subsection{Decays of pseudoscalar mesons}
\label{sec:generic_P_decay_section}

\begin{figure}[ht]
\begin{center}
\includegraphics[width=8cm]{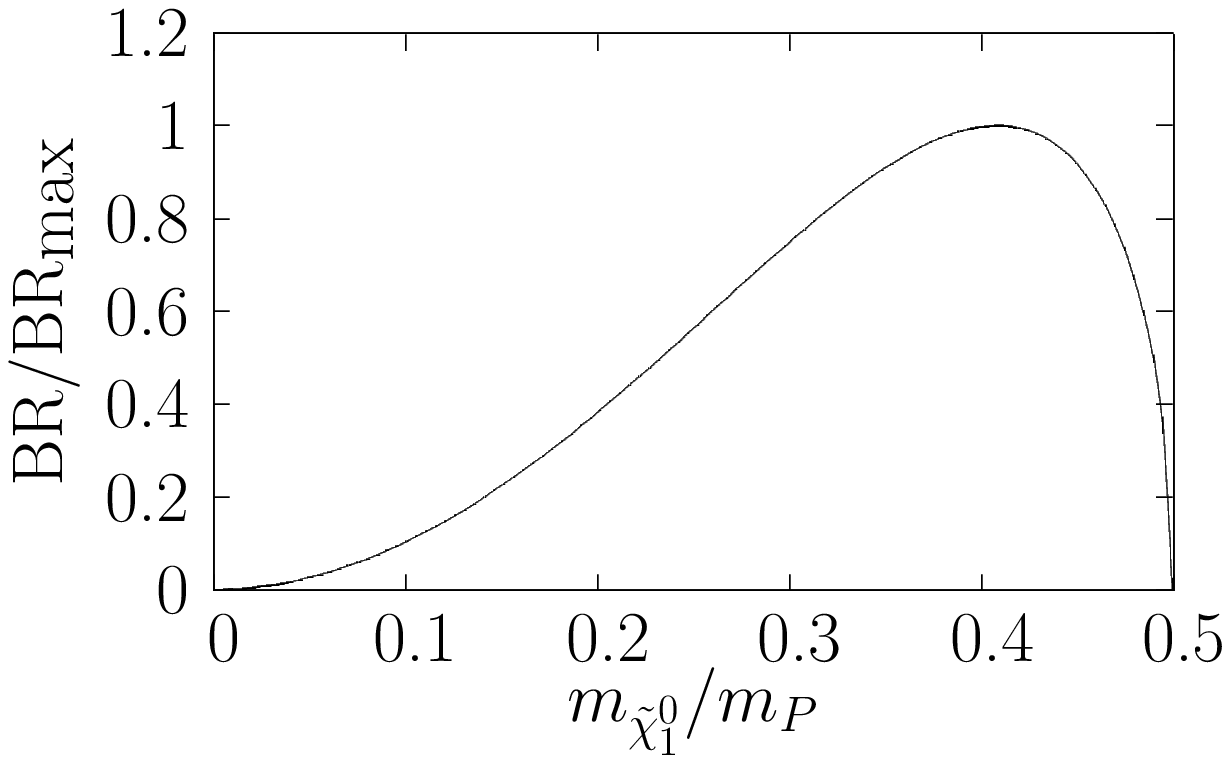}
\caption{\label{fig:BRpseudoscalar} The branching ratio for a
  pseudoscalar meson decay $P {\to} {\xone} {\xone}$ against
  the ratio ${\mxone} / m_{P}$ normalized to the peak branching
  ratio.}
\end{center}
\end{figure}

We first consider the case ${\pi}^{0} {\to} {\xone} {\xone}$, and then
generalize to the other pseudoscalar meson decays.  Since the
branching ratio depends on the ratio ${\mxone} / m_{P}$ in exactly the
same way for each pseudoscalar meson $P$ and differs only by an
overall factor, we plot the dependence once, normalized to the peak
branching ratio, in Fig.~\ref{fig:BRpseudoscalar}. Throughout this
paper, to avoid unnecessary clutter, we remove the superscripts
denoting charges for mesons in subscripts, \eg~we write
$m_{{\pi}}$ for the mass of the neutral pion, rather than
$m_{{{\pi}^{0}}}$, and we ignore the differences in masses between
charged and uncharged mesons, so $m_{{\pi}}$ is also used for the mass
of the charged pion.

The decay of a pseudoscalar meson into a
fermion-antifermion pair requires at least one of the fermions to be
massive, due to the required helicity flip.  In
Fig.~\ref{fig:BRpseudoscalar} we see that the branching ratio goes to
zero for vanishing neutralino mass.  Hence pseudoscalar decays offer no
bounds for models with massless neutralinos.  Instead, the peak
branching ratio is reached for
${\mxone} / m_{P} = 1 / {\sqrt{6}} \approx 0.41$.

Likewise, the SM background of $P {\to} {\nu}_i {\bar{{\nu}}}_i$, $i =
e, {\mu}, {\tau}$, is extremely small.  Its branching ratio is
proportional to the neutrino mass squared divided by the pion mass
squared.  Even for the pion and relatively heavy neutrinos with
$m_{{{\nu}_{i}}} = 1$ eV, this is of the order of $10^{-10}$.

\subsubsection{${\pi}^{0} {\to} {\xone} {\xone}$}

We treat the neutral pion as a bound state of valence quarks, given in
terms of quark field bilinears as
$( u {\bar{u}} - d {\bar{d}} ) / {\sqrt{2}}$.  Denoting the neutralino
momenta by $k_{1}$ and $k_{2}$, the matrix element for this decay is
given by
\begin{eqnarray}
{\mathcal{M}} & = & {\frac{1}{i}} {\bra{{\xone} {\xone}}}
\left[ {\xbarone} ( i |e| a_{u} {\superpartner{u}}_{L}^{{\ast}} P_{L}
+ i |e| b_{u} {\superpartner{u}}_{R}^{{\ast}} P_{R} ) u
+ {\xbarone} ( i |e| a_{d} {\superpartner{d}}_{L}^{{\ast}} P_{L}
+ i |e| b_{d} {\superpartner{d}}_{R}^{{\ast}} P_{R} ) d \right] \nonumber\\
 & & {\times} \left[ {\bar{u}} ( -i |e| a_{u} {\superpartner{u}}_{L} P_{R} 
- i |e| b_{u} {\superpartner{u}}_{R} P_{L} ) {\xone}
+ {\bar{d}} ( -i |e| a_{d} {\superpartner{d}}_{L} P_{R}
- |e| i b_{d} {\superpartner{d}}_{R} P_{L} ) {\xone} \right]
{\ket{{{\pi}^{0}}}} \nonumber\\
 & = & {\frac{{e^{2}}}{8}} {\bar{u}}_{{\xone}}( k_{1} ) {\gamma}_{{\mu}} 
{\gammabar} v_{{\xone}}( k_{2} )\nonumber \\
&& \times\; {\bra{0}}
\left( {\frac{{a_{u}^{2}}}{{m_{{\superpartner{u}}_{L}}^{2}}}}
+ {\frac{{b_{u}^{2}}}{{m_{{\superpartner{u}}_{R}}^{2}}}} \right) 
{\bar{u}} {\gamma}^{{\mu}} {\gammabar} u
+ \left( {\frac{{a_{d}^{2}}}{{m_{{\superpartner{d}}_{L}}^{2}}}}
+ {\frac{{b_{d}^{2}}}{{m_{{\superpartner{d}}_{R}}^{2}}}} \right)
{\bar{d}} {\gamma}^{{\mu}} {\gammabar} d {\ket{{{\pi}^{0}}}}
- ( k_{1} {\leftrightarrow} k_{2} ). 
\label{pion-matrix-el}
\end{eqnarray}
In the last line ${\bar{u}}_{{\xone}}( k_{1} )$ and
$v_{{\xone}}( k_{2} )$ denote the standard $4$-component Dirac
wave functions (for the neutralinos), whereas
$u, {\bar{u}}, d, {\bar{d}}$ refer to the corresponding quarks.
We have approximated the squark propagators by
$-i / m_{{\superpartner{q}}_{L/R}}^{2}$, since the pion mass $m_{{\pi}}$
(and in general all the pseudoscalar masses $m_{P}$)
${\ll} m_{{\superpartner{q}}}$.  In order to obtain the last line
we made use of the fact that the pion is a pseudoscalar state,
\ie~${\bra{0}} {\bar{u}} {\gamma}^{{\mu}} u {\ket{{{\pi}^{0}}}}
= {\bra{0}} {\bar{d}} {\gamma}^{{\mu}} d {\ket{{{\pi}^{0}}}} = 0$.
Since the neutralino is a Majorana fermion, we also have ${\xbarone} {\gamma}^{{\mu}} {\xone} = {\xbaroneL} {\sigma}^{{\mu}} {\xoneL} + {\xoneL} {\bar{{\sigma}}}^{{\mu}} {\xbaroneL} = 0$ identically.  We used Fierz
identities~\cite{Bailin:1994qt,Dreiner:2008tw} to transform the
products of spinor bilinears to the presented form.

For purely bino neutralinos, the neutralino-quark-squark couplings are
$|e| a_{q} P_{L}$ for an outgoing ${\superpartner{q}}_{L}$ and
$|e| b_{q} P_{R}$ for an outgoing ${\superpartner{q}}_{R}$, and since
the gauge couplings are generation-independent, we just write $a_{u}$ for
up-type quarks and $a_{d}$ for down-type quarks, with
\begin{equation}
a_{u} = a_{d} = {\frac{{-1}}{3 {\sqrt{2}} {\cos} {\theta}_{W} }},\quad
b_{u} = {\frac{{2 {\sqrt{2}}}}{{3 {\cos} {\theta}_{W} }}}, \quad
b_{d} = {\frac{{-{\sqrt{2}}}}{{3 {\cos} {\theta}_{W} }}}\,.
\end{equation}
$e$ is the electromagnetic charge of the electron, such that $e^{2} =
4 {\pi} {\alpha}$.  We take ${\alpha} = 1 / 137$ and the electroweak
mixing angle $\sin^2\theta_W = 0.23$.  Eq.~(\ref{pion-matrix-el}) can
be seen to agree with Eqs.~($2$) and ($3$) in
Ref.~\cite{Dobroliubov:1987cba}, if the neutralino-quark-squark
couplings are chosen to be photino-like instead of bino-like [\ie\
$a_{u} = b_{u} = 2 {\sqrt{2}} / 3$ and $a_{d} = b_{d} = -{\sqrt{2}} /
3$] and the squark mixing ${\mu}^{2}$ therein is set to zero.

We use the standard definition of the pseudoscalar decay
constant~\cite{oldPDG, Leutwyler} to write
\begin{equation}\label{eq:fpi}
{\bra{0}} {\bar{u}} {\gamma}^{{\mu}} {\gammabar} u {\ket{{{\pi}^{0}}}} 
= -{\bra{0}} {\bar{d}} {\gamma}^{{\mu}} {\gammabar} d {\ket{{{\pi}^{0}}}} 
= {\frac{i}{{\sqrt{2}}}} F_{{\pi}} p_{{\pi}}^{{\mu}},
\end{equation}
where $p_{{\pi}}$ is the momentum of the pion. Eq.~(\ref{eq:fpi})
defines our convention for the value of the pion decay constant
$F_{{\pi}}$ as $131$~MeV~\cite{PDG} (as opposed to the other common
convention of $91$ MeV, which incorporates the factor of
$1 / {\sqrt{2}}$).

Summing over final spins, the square of the matrix element is given by
\begin{equation}
{\overline{{| {\mathcal{M}} |^{2}}}} =
{\frac{{e^{4} F_{{\pi}}^{2} {\mxonesq}}} {{8 m_{{\pi}}^{2}}}}
( 2 {\dotproduct{{p_{{\pi}}}}{{k_{1}}}} {\dotproduct{{p_{{\pi}}}}{{k_{2}}}}
- m_{{\pi}}^{2} {\dotproduct{{k_{1}}}{{k_{2}}}}
+ m_{{\pi}}^{2} {\mxonesq} )
\left( {\frac{{a_{u}^{2}}}{{m_{{\superpartner{u}}_{L}}^{2}}}} +
{\frac{{b_{u}^{2}}}{{m_{{\superpartner{u}}_{R}}^{2}}}}
- {\frac{{a_{d}^{2}}}{{m_{{\superpartner{d}}_{L}}^{2}}}}
- {\frac{{b_{d}^{2}}}{{m_{{\superpartner{d}}_{R}}^{2}}}} \right)^{2},
\end{equation}
leading to a partial decay width of
\begin{equation}
{\Gamma}_{{{\pi}^{0} {\to} {\xone} {\xone}}} =
{\frac{{{\alpha}^{2} {\pi}}}{8}} F_{{\pi}}^{2} {\mxonesq} m_{{\pi}}
{\sqrt{{1 - 4 \left( {\frac{{\mxone}}{{m_{{\pi}}}}} \right)^{2}}}}
\left( {\frac{{a_{u}^{2}}}{{m_{{\superpartner{u}}_{L}}^{2}}}}
+ {\frac{{b_{u}^{2}}}{{m_{{\superpartner{u}}_{R}}^{2}}}}
- {\frac{{a_{d}^{2}}}{{m_{{\superpartner{d}}_{L}}^{2}}}}
- {\frac{{b_{d}^{2}}}{{m_{{\superpartner{d}}_{R}}^{2}}}} \right)^{2}.
\end{equation}

If we assume a common squark mass $m_{{\superpartner{q}}}$, then we
obtain a branching ratio of
\begin{equation}
  \mbox{BR}( {\pi}^{0} {\to} {\xone}{\xone})=(4.64 {\times} 10^{-14}) 
  \left( {\frac{{100 \mbox{ GeV}}}{{m_{{\superpartner{q}}}}}} \right)^{4} 
  \left( {\frac{{\mxone}}{{1 \mbox{ MeV}}}} \right)^{2}
   {\sqrt{{1 - ( 2.20 {\times}  10^{-4} )
   \left( {\frac{{\mxone}}{{1 \mbox{ MeV}}}} \right)^{2}}}}, 
\end{equation}
using $m_{{\pi}} = 135$ MeV and
$\Gamma_{{\pi}^{0}, \mathrm{tot}} = 7.8$~eV~\cite{PDG}.

Since there is destructive interference between the up and down quark
contributions, we consider an extreme case where the down-type squarks
are so heavy that they decouple.  This maximizes the branching ratio to
\begin{equation}
\mbox{BR}( {\pi}^{0} {\to} {\xone} {\xone} ) =
(9.30 {\times} 10^{-14})
\left( {\frac{{100 \mbox{ GeV}}}{{m_{{\superpartner{u}}}}}} \right)^{4}
\left( {\frac{{\mxone}}{{1 \mbox{ MeV}}}} \right)^{2}
{\sqrt{{1 - ( 2.20 {\times}  10^{-4})
\left( {\frac{{\mxone}}{{1 \mbox{ MeV}}}} \right)^{2}}}}.
\end{equation}

Setting the up squark masses to $100$~GeV, the maximal branching ratio
is $1.63 {\times} 10^{{-10}}$, when ${\mxone} = 55.1$~MeV.  The
current experimental upper bound on the decay of the neutral pion to
invisible particles is $2.7 {\times} 10^{-7}$ \cite{Artamonov:2005cu},
which is at least three orders of magnitude greater.  Hence we do not
find significant constraints on supersymmetric models with extremely
light bino-like neutralinos from pion decays.

\subsubsection{${\eta} {\to} {\xone} {\xone}$, ${{\eta}'}{\to} {\xone} {\xone}$}

We consider the decays of ${\eta}$ and ${{\eta}'}$ mesons analogously.
We take into account mixing between the ${\eta}^{0}$ and ${\eta}^{8}$
$SU(3)$-flavor states, as described by
Ref.~\cite{Feldmann:1998vh}, which leads to:
\begin{eqnarray}
{\bra{0}} {\bar{u}} {\gamma}^{{\mu}} {\gammabar} u {\ket{{\eta}}} = 
{\bra{0}} {\bar{d}} {\gamma}^{{\mu}} {\gammabar} d {\ket{{\eta}}} 
& = & i \left( {\frac{1}{{\sqrt{6}}}} {\cos} {\theta}_{8}  
F_{{{\eta}^{8}}} - {\frac{1}{{\sqrt{3}}}} {\sin}{\theta}_{1}  
F_{{{\eta}^{1}}} \right) p_{{\eta}}^{{\mu}} \equiv i {\tilde{F}}_
{{{\bar{u}}u/{\bar{d}}d}}^{{\eta}} p_{{\eta}}^{{\mu}} \nonumber \\
{\bra{0}} {\bar{s}} {\gamma}^{{\mu}} {\gammabar} s {\ket{{\eta}}} 
& = & i \left( {\frac{{-2}}{{\sqrt{6}}}} {\cos} {\theta}_{8}  
F_{{{\eta}^{8}}} - {\frac{1}{{\sqrt{3}}}} {\sin}{\theta}_{1}  
F_{{{\eta}^{1}}} \right) p_{{\eta}}^{{\mu}} \equiv i {\tilde{F}}_
{{{\bar{s}}s}}^{{\eta}} p_{{\eta}}^{{\mu}}
\end{eqnarray}
and
\begin{eqnarray}
{\bra{0}} {\bar{u}} {\gamma}^{{\mu}} {\gammabar} u {\ket{{{\eta}'}}} 
= {\bra{0}} {\bar{d}} {\gamma}^{{\mu}} {\gammabar} d {\ket{{{\eta}'}}} 
& = & i \left( {\frac{1}{{\sqrt{6}}}} {\sin}{\theta}_{8}  
F_{{{\eta}^{8}}} + {\frac{1}{{\sqrt{3}}}} {\cos} {\theta}_{1}  
F_{{{\eta}^{1}}} \right) p_{{{\eta}'}}^{{\mu}} \equiv
i {\tilde{F}}_{{{\bar{u}}u/{\bar{d}}d}}^{{{\eta}'}} p_{{{\eta}'}}^{{\mu}}
\nonumber \\
{\bra{0}} {\bar{s}} {\gamma}^{{\mu}} {\gammabar} s {\ket{{{\eta}'}}} 
& = & i \left( {\frac{{-2}}{{\sqrt{6}}}} {\sin}{\theta}_{8} 
F_{{{\eta}^{8}}} + {\frac{1}{{\sqrt{3}}}} {\cos} {\theta}_{1}  
F_{{{\eta}^{1}}} \right) p_{{{\eta}'}}^{{\mu}} \equiv
i {\tilde{F}}_{{{\bar{s}}s}}^{{{\eta}'}} p_{{{\eta}'}}^{{\mu}}\,.
\end{eqnarray}
The phenomenological values of the decay constants and mixing angles
are given in Table~\ref{tab:eta1_8_mixing}.

\begin{table}
\begin{center}
\begin{tabular}{c d@{\hspace{1mm}} l} 
\hline \\[-9mm] \hline
\vspace{0.1cm} ${\theta}_{8}$  & 0.37 & rad \\
\vspace{0.1cm} ${\theta}_{1}$ & 0.16 & rad \\
\vspace{0.1cm} $F_{{{\eta}^{8}}} / F_{{\pi}}$ & 1.28 & \\
\vspace{0.1cm} $F_{{{\eta}^{1}}} / F_{{\pi}}$ & 1.17 & \\ \hline \\[-9mm] \hline
\end{tabular}
\end{center}
\caption{Input parameters for ${\eta}$ and ${{\eta}'}$ decay
  \cite{Feldmann:1998vh}.} \label{tab:eta1_8_mixing}
\end{table}

This leads to a partial decay width of
\begin{eqnarray}
{\Gamma}_{{{\eta}^{{( {\prime} )}} {\to} {\xone} {\xone}}}
 & = & 
{\frac{{{\alpha}^{2} {\pi}}}{4}} {\mxonesq}
m_{{{\eta}^{{( {\prime} )}}}}
{\sqrt{{1 - 4 \left(
{\frac{{\mxone}}{{m_{{{\eta}^{{( {\prime} )}}}}}}} \right)^{2}}}}
\nonumber\\
 & & {\times}
\left[ {\tilde{F}}_{{{\bar{u}}u/{\bar{d}}d}}^{{{\eta}^{{( {\prime} )}}}}
\left( {\frac{{a_{u}^{2}}}{{m_{{\superpartner{u}}_{L}}^{2}}}} 
+ {\frac{{b_{u}^{2}}}{{m_{{\superpartner{u}}_{R}}^{2}}}}
+ {\frac{{a_{d}^{2}}}{{m_{{\superpartner{d}}_{L}}^{2}}}}
+ {\frac{{b_{d}^{2}}}{{m_{{\superpartner{d}}_{R}}^{2}}}} \right)
+ {\tilde{F}}_{{{\bar{s}}s}}^{{{\eta}^{{( {\prime} )}}}}
\left( {\frac{{a_{d}^{2}}}{{m_{{\superpartner{s}}_{L}}^{2}}}}
+ {\frac{{b_{d}^{2}}}{{m_{{\superpartner{s}}_{R}}^{2}}}} \right) \right]^{2}.
\end{eqnarray}

If we again assume a common squark mass $m_{{\superpartner{q}}}$, then
we obtain the branching ratios:
\begin{equation}
\mbox{BR}( {\eta} {\to} {\xone} {\xone} ) = ( 1.18 {\times} 10^{-15} ) 
\left( {\frac{{100 \mbox{ GeV}}}{{m_{{\superpartner{q}}}}}} \right)^{4} 
\left( {\frac{{\mxone}}{{1 \mbox{ MeV}}}} \right)^{2}
{\sqrt{{1 - ( 1.33 {\times}  10^{-5} )
\left( {\frac{{\mxone}}{{1 \mbox{ MeV}}}} \right)^{2}}}}
\end{equation}
and
\begin{equation}
\mbox{BR}( {{\eta}'} {\to} {\xone} {\xone} ) = ( 4.34 {\times} 10^{-17} ) 
\left( {\frac{{100 \mbox{ GeV}}}{{m_{{\superpartner{q}}}}}} \right)^{4} 
\left( {\frac{{\mxone}}{{1 \mbox{ MeV}}}} \right)^{2}
{\sqrt{{1 - ( 4.36 {\times}  10^{-5} )
\left( {\frac{{\mxone}}{{1 \mbox{ MeV}}}} \right)^{2}}}}\,.
\label{etaprimeBReqn}
\end{equation}
If we decouple the strange squarks to maximize the branching ratio, we
obtain for the ${\eta}$
\begin{equation}
\mbox{BR}( {\eta} {\to} {\xone} {\xone} ) = ( 2.63 {\times} 10^{-15} ) 
\left( {\frac{{100 \mbox{ GeV}}}{{m_{{\superpartner{q}}}}}} \right)^{4} 
\left( {\frac{{\mxone}}{{1 \mbox{ MeV}}}} \right)^{2}
{\sqrt{{1 - ( 1.33 {\times}  10^{-5} )
\left( {\frac{{\mxone}}{{1 \mbox{ MeV}}}} \right)^{2}}}}.
\end{equation}
We use $m_{{\eta}} = 548$ MeV, $m_{{{\eta}'}} = 958$ MeV,
$\Gamma_{{\eta}, \mathrm{tot}} = 1.30$ keV and
$\Gamma_{{{\eta}'},\mathrm{tot}} = 205$ keV, respectively~\cite{PDG}.

Since in the case of the ${{\eta}'}$ decay the contributions 
from each flavor of quark constructively interfere, we use the 
expression with a common squark mass, Eq.~(\ref{etaprimeBReqn}), to 
obtain a maximal branching ratio.  Setting the non-decoupled squark 
masses to $100$ GeV, the maximal branching ratios are
$7.60 {\times} 10^{-11}$ for ${\eta} {\to} {\xone} {\xone}$ and
$3.83 {\times} 10^{{-12}}$ for ${{\eta}'} {\to} {\xone} {\xone}$,
when ${\mxone} = 223$ MeV and $391$ MeV, respectively.  The current
experimental upper bound on the decay of the ${\eta}$ to invisible
particles is $6 {\times} 10^{-4}$ \cite{PDG}, and for the
${{\eta}'}$ it is $1.4 {\times} 10^{-3}$ \cite{PDG}.  Thus
for ${\eta}$ and ${{\eta}'}$ decays we also
do not find significant constraints on models with light
bino-like neutralinos.

\subsection{Decays of vector mesons}

\begin{figure}[t]
\begin{center}
\includegraphics[width=8cm]{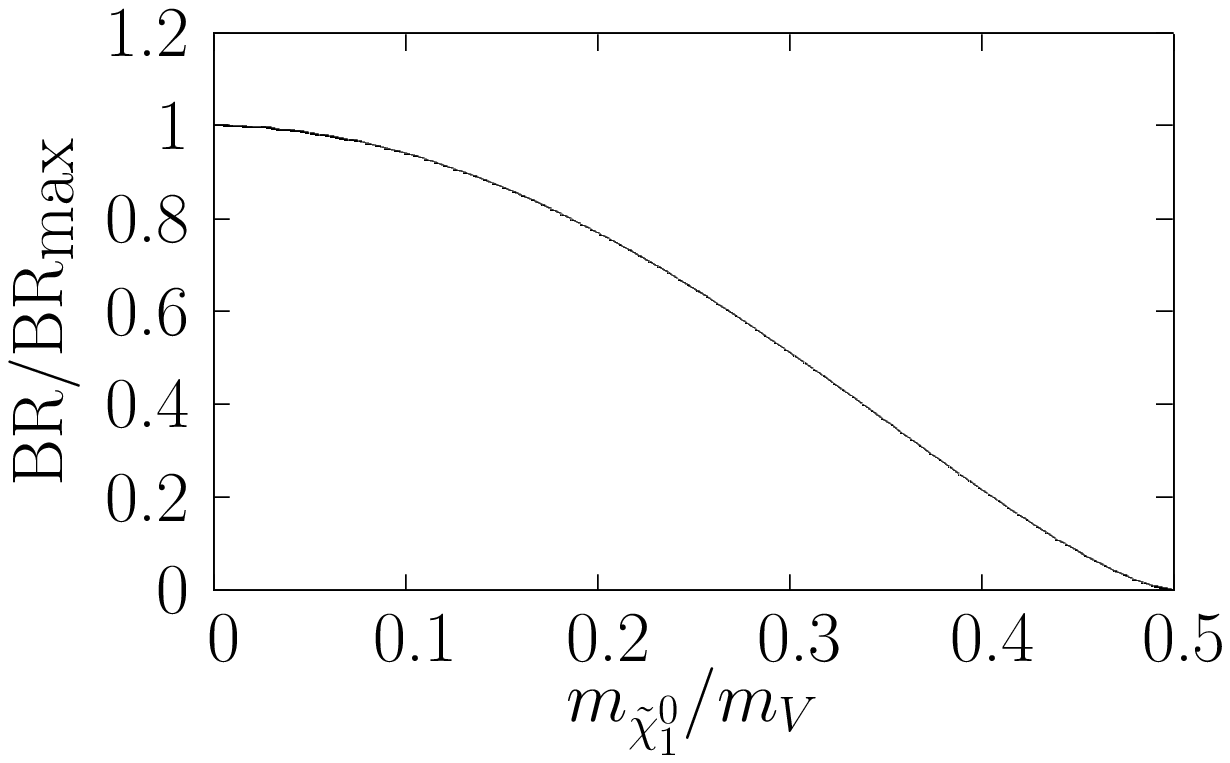}
\caption{\label{fig:BRvector} The branching ratio for a vector meson
  decay $V {\to} {\xone} {\xone}$ against the ratio ${\mxone} / m_{V}$
  normalized to the peak branching ratio.}
\end{center}
\end{figure}

We first consider the case of ${\rho}^{0} {\to} {\xone} {\xone}$, and
then generalize to the other vector meson decays.  Since the branching
ratio depends on the ratio ${\mxone} / m_{V}$ in exactly the same way
for each vector meson $V$ and differs only by an overall factor, we
plot the dependence once, normalized to the peak branching ratio, in
Fig.~\ref{fig:BRvector}.  We note that there is no helicity suppression
in this case, and so increasing the ${\xone}$ mass only reduces the
available phase space and thus the branching ratio.

Note that ${}^3S_1$ quarkonium states can only decay into gauginos if
parity is violated.  As photinos and gluinos couple to left- and
right-handed squarks with equal strength, the decays 
${}^3S_1(q\bar{q}) \to \tilde{\gamma}\tilde{\gamma}$ and 
${}^3S_1(q\bar{q}) \to \tilde{g}\tilde{g}$ can only proceed if parity 
violation is introduced by a difference in the left- and right-handed
squark masses~\cite{Ellis:1983ed}. 
Binos, however, couple more strongly to right-handed
particles than to left-handed particles, \ie~parity is
explicitly violated, and quarkonium decays into binos are possible
also for degenerate left- and right-handed squark masses.

\subsubsection{${\rho}^{0} {\to} {\xone} {\xone}$}

We treat the ${\rho}^{0}$ as the vector equivalent of the neutral pion
[\ie\ the same valence quarks,
$( u {\bar{u}} - d {\bar{d}} ) / {\sqrt{2}} $].  The calculation of
the matrix element is similar, except we use
\begin{equation}
{\bra{0}} {\bar{u}} {\gamma}^{{\mu}} {\gammabar} u {\ket{{{\rho}^{0}}}} 
= {\bra{0}} {\bar{d}} {\gamma}^{{\mu}} {\gammabar} d {\ket{{{\rho}^{0}}}} = 0
\end{equation}
due to the CP-properties of the ${\rho}^{0}$ meson, and we define
\begin{equation}
{\bra{0}} {\bar{u}} {\gamma}^{{\mu}} u {\ket{{{\rho}^{0}}}} =
-{\bra{0}} {\bar{d}} {\gamma}^{{\mu}} d {\ket{{{\rho}^{0}}}} =
{\frac{i}{{\sqrt{2}}}} m_{{\rho}} F_{{\rho}} {\epsilon}_{{\rho}}^{{\mu}},
\label{vector_fudge_equation}
\end{equation}
where ${\epsilon}_{{\rho}}^{{\mu}}$ is the polarization vector of the
${\rho}^{0}$.  The factor $F_{{\rho}}$ we obtain by calculating the SM
decay width of ${\rho}^{0} {\to} e^{+} e^{-}$ in this approximation to
leading order in ${\alpha}$:
\begin{equation}
\label{eq:rho_width}
{\Gamma}_{{{\rho}^{0} {\to} e^{+} e^{-}}} =
{\frac{{4 {\pi} {\alpha}^{2}}}{3}}
{\frac{{F_{{\rho}}^{2}}}{{m_{{\rho}}}}}
\left( Q_{u} H_{u} + Q_{d} H_{d} \right)^{2} =
{\frac{{2 {\pi} {\alpha}^{2}}}{3}}
{\frac{{F_{{\rho}}^{2}}}{{m_{{\rho}}}}}\,,
\end{equation}
where $Q_{u} = 2 / 3$, $Q_{d} = -1 / 3$ (the quark charges in units of
$|e|$) and $H_{u} = 1 / {\sqrt{2}}$, $H_{d} = -1 / {\sqrt{2}}$.
$H_{u,d}$ are the coefficients of the $u {\bar{u}}$
and $d {\bar{d}}$ quark field bilinears in the quark model wave
function, \eg\ ${\rho}^{0} = ( u {\bar{u}} - d {\bar{d}} )/ {\sqrt{2}}
{\equiv} ( H^{\rho}_u u {\bar{u}} + H^{\rho}_d d {\bar{d}})$, see
Ref.~\cite{Dreiner:2006gu} for further details.  With this generic
notation, Eq.~(\ref{eq:rho_width}) can easily be adjusted for other
vector mesons, \eg\ for ${\Upsilon}$, $H_{b} = 1$, no other $H$,
$Q_{b} = -1 / 3$, so
${\Gamma}_{{{\Upsilon}{\to} e^{+} e^{-}}}=
(4{\pi}{\alpha}^{2}/27)(F_{{\Upsilon}}^{2} / m_{{\Upsilon}})$
\footnote{Note that the decay of heavy-quark bound states like
  $J/{\psi}$ or ${\Upsilon}$ can be treated systematically in
  non-relativistic QCD, see \eg\ Ref.~\cite{Brambilla:2004wf}.}.

Thus for ${\rho}^{0} {\to} {\xone} {\xone}$ we obtain for the matrix
element
\begin{equation}
\mathcal{M} = {\frac{{e^{2}}}{8}}
{\bar{u}}_{\tilde\chi^0_1}( k_{1} ) {\gamma}_{{\mu}}
{\gammabar} v_{\tilde\chi^0_1}( k_{2} ) {\bra{0}}
\left( {\frac{{b_{u}^{2}}}{{m_{{\superpartner{u}}_{R}}^{2}}}}
- {\frac{{a_{u}^{2}}}{{m_{{\superpartner{u}}_{L}}^{2}}}} \right)
{\bar{u}} {\gamma}^{{\mu}} u
+ \left( {\frac{{b_{d}^{2}}}{{m_{{\superpartner{d}}_{R}}^{2}}}}
- {\frac{{a_{d}^{2}}}{{m_{{\superpartner{d}}_{L}}^{2}}}} \right)
{\bar{d}} {\gamma}^{{\mu}} d {\ket{{{\rho}^{0}}}}
- ( k_{1} {\leftrightarrow} k_{2} )\,.
\end{equation}
After squaring, averaging over initial spins, and summing over final
spins, we obtain
\begin{equation}
{\overline{{| {\mathcal{M}} |^{2}}}} =
{\frac{{e^{4} F_{{\rho}}^{2}}}{{96}}} 
\left[ \left( {\frac{{b_{u}^{2}}}{{m_{{\superpartner{u}}_{L}}^{2}}}} 
- {\frac{{a_{u}^{2}}}{{m_{{\superpartner{u}}_{R}}^{2}}}} \right)
- \left( {\frac{{b_{d}^{2}}}{{m_{{\superpartner{d}}_{L}}^{2}}}}
- {\frac{{a_{d}^{2}}}{{m_{{\superpartner{d}}_{R}}^{2}}}} \right) \right]^{2}
( 2 {\dotproduct{{k_{1}}}{{p_{{\rho}}}}}
{\dotproduct{{k_{2}}}{{p_{{\rho}}}}}
+ m_{{\rho}}^{2} {\dotproduct{{k_{1}}}{{k_{2}}}}
- 3 m_{{\rho}}^{2} {\mxonesq} ),
\end{equation}
leading to the partial decay width 
\begin{eqnarray}
{\Gamma}_{{{\rho}^{0} {\to} {\xone} {\xone}}} & = &
{\frac{{{\alpha}^{2} {\pi}}}{192}} F_{{\rho}}^{2} m_{{\rho}}^{3}
\left[ 1 - 4
\left( {\frac{{\mxone}}{{m_{{\rho}}}}} \right)^{2} \right]^{{( 3 / 2 )}} 
\left( {\frac{{b_{u}^{2}}}{{m_{{\superpartner{u}}_{R}}^{2}}}}
- {\frac{{a_{u}^{2}}}{{m_{{\superpartner{u}}_{L}}^{2}}}}
- {\frac{{b_{d}^{2}}}{{m_{{\superpartner{d}}_{R}}^{2}}}}
+ {\frac{{a_{d}^{2}}}{{m_{{\superpartner{d}}_{L}}^{2}}}} \right)^{2} 
\nonumber\\[3mm]
 & = & {\frac{1}{{32}}}
{\Gamma}_{{{\rho}^{0} {\to} e^{+} e^{-}}} m_{{\rho}}^{4}
\left[ 1 - 4
\left( {\frac{{\mxone}}{{m_{{\rho}}}}} \right)^{2} \right]^{{( 3 / 2 )}}
\left( {\frac{{b_{u}^{2}}}{{m_{{\superpartner{u}}_{R}}^{2}}}}
- {\frac{{a_{u}^{2}}}{{m_{{\superpartner{u}}_{L}}^{2}}}}
- {\frac{{b_{d}^{2}}}{{m_{{\superpartner{d}}_{R}}^{2}}}}
+ {\frac{{a_{d}^{2}}}{{m_{{\superpartner{d}}_{L}}^{2}}}} \right)^{2}.
\end{eqnarray}

If we minimize destructive interference by decoupling the
left-handed up squark and the right-handed down squark, we obtain
\begin{equation}
\mbox{BR}( {\rho}^{0} {\to} {\xone} {\xone} ) = ( 8.01 {\times} 10^{-15} ) 
\left( {\frac{{100 \mbox{ GeV}}}{{m_{{\superpartner{q}}}}}} \right)^{4} 
\left[ 1 - ( 6.65 {\times} 10^{-6} )
\left( {\frac{{\mxone}}{{1 \mbox{ MeV}}}} \right)^{2} \right]^{{( 3 / 2 )}},
\end{equation}
using $m_{\rho} = 775$ MeV,
${\Gamma}_{{{\rho}^{0} {\to} e^{+} e^{-}}} / \Gamma_{{\rho}^{0}, \mathrm{tot}}
= 4.71 {\times} 10^{-5}$~\cite{PDG}.  
The branching ratio is maximized for ${\mxone} = 0$.  Setting the
non-decoupled squark masses to $100$ GeV, we find a branching ratio of
$8.01 {\times} 10^{-15}$.  We can find no experimental bound at this
time for the branching ratio of ${\rho}^{0} {\to}$ invisible, but note
that a branching ratio of $10^{-14}$ is far below the bounds on other
branching ratios measured for the ${\rho}^{0}$ meson \cite{PDG}.

\subsubsection{${\omega} {\to} {\xone} {\xone}$, ${\phi} {\to} {\xone} {\xone}$}

The mixing between the $SU(3)$-flavor states amongst the vector mesons
is very different from the pseudoscalars.  Conveniently, it is such
that the ${\phi}$ meson is nearly pure $s{\bar{s}}$ and the ${\omega}$
is $( u {\bar{u}} + d {\bar{d}} ) / {\sqrt{2}}$, and we treat them as
these pure states.

This means that the decay ${\omega} {\to} {\xone} {\xone}$ is analogous
to ${\rho} {\to} {\xone} {\xone}$, taking into account the constructive
rather than destructive interference between the up and down quarks:
\begin{equation}
{\Gamma}_{{{\omega} {\to} {\xone} {\xone}}} = {\frac{9}{32}} 
{\Gamma}_{{{\omega} {\to} e^{+} e^{-}}} m_{{\omega}}^{4}
\left[ 1 - 4
\left( {\frac{{\mxone}}{{m_{{\omega}}}}} \right)^{2} \right]^{{( 3 / 2 )}}
\left( {\frac{{b_{u}^{2}}}{{m_{{\superpartner{u}}_{R}}^{2}}}}
- {\frac{{a_{u}^{2}}}{{m_{{\superpartner{u}}_{L}}^{2}}}}
+ {\frac{{b_{d}^{2}}}{{m_{{\superpartner{d}}_{R}}^{2}}}}
- {\frac{{a_{d}^{2}}}{{m_{{\superpartner{d}}_{L}}^{2}}}} \right)^{2}.
\end{equation}

If we take a conservative estimate by decoupling the left-handed up
and down squark masses, we obtain 
\begin{equation}
\mbox{BR}( {\omega} {\to} {\xone} {\xone} ) =
( 1.57 \times 10^{-13} )
\left({\frac{{100 \mbox{ GeV}}}{{m_{{{\superpartner{q}}_{R}}}}}} \right)^{4} 
\left[ 1 - ( 6.53 {\times} 10^{-6} )
\left( {\frac{{\mxone}}{{1 \mbox{ MeV}}}} \right)^{2} \right]^{{( 3 / 2 )}}, 
\end{equation}
with $m_{\omega} = 783$ MeV,
${\Gamma}_{{{\omega} {\to} e^{+} e^{-}}} / \Gamma_{{\omega}, \mathrm{tot}}
= 7.16 {\times} 10^{-5}$~\cite{PDG}.  
This branching ratio is maximized for ${\mxone} = 0$.  Setting the
non-decoupled squark masses to $100$ GeV, we find
BR$( {\omega} {\to} {\xone} {\xone} ) = 1.57 {\times} 10^{-13}$.

The ${\phi}$ case is particularly simple.  The partial width is given
by
\begin{equation}
{\Gamma}_{{{\phi} {\to} {\xone} {\xone}}} =
{\frac{9}{32}} {\Gamma}_{{{\phi} {\to} e^{+} e^{-}}} m_{{\phi}}^{4}
\left[ 1 - 4
\left( {\frac{{\mxone}}{{m_{{\phi}}}}} \right)^{2} \right]^{{( 3 / 2 )}}
\left( {\frac{{b_{d}^{2}}}{{m_{{\superpartner{s}}_{R}}^{2}}}}
- {\frac{{a_{d}^{2}}}{{m_{{\superpartner{s}}_{L}}^{2}}}} \right)^{2}.
\end{equation}

If we minimize destructive interference by decoupling the
left-handed strange squark mass, we obtain 
\begin{equation}
\mbox{BR}( {\phi} {\to} {\xone} {\xone} ) = ( 7.51 \times 10^{-14} )
\left({\frac{{100\mbox{ GeV}}}{{m_{{{\superpartner{s}}_{R}}}}}} \right)^{4}
\left[ 1 - ( 3.85 {\times} 10^{-6} )
\left( {\frac{{\mxone}}{{1 \mbox{ MeV}}}} \right)^{2} \right]^{{( 3 / 2 )}},
\end{equation}
which is maximized for ${\mxone} = 0$.  Here $m_{\phi} = 1.02$ GeV,
${\Gamma}_{{{\phi} {\to} e^{+} e^{-}}} / \Gamma_{{\phi}, \mathrm{tot}} =
2.97 {\times} 10^{-4}$~\cite{PDG}.  Setting the right-handed strange 
squark mass to $100$ GeV, we find a branching ratio of 
$7.51 {\times} 10^{-14}$.  Again, we can find no experimental bound 
at this time for the branching ratios of ${\phi}$ or ${\omega} {\to}$ 
invisible, but note that branching ratios of $10^{-13}$ and $10^{-14}$
are far below the bounds on other branching ratios measured for these 
mesons \cite{PDG}.

\subsubsection{$J/{\psi} {\to} {\xone} {\xone}$, ${\Upsilon} {\to} {\xone} {\xone}$}

The squarks associated with the heavier flavors of quarks (charm,
bottom and top) are expected to mix their left-handed and right-handed
components much more than the other flavors \cite{susy2,susy3}. We take
this into account for the $J/{\psi}$ and ${\Upsilon}$ decays. 
Writing the mass eigenstates of the charm squarks
${\superpartner{c}}_{1,2}$ as
\begin{equation}
\left({\myatop{{{\superpartner{c}}_1}}{{{\superpartner{c}}_2}}}\right) = 
\left( \begin{array}{ c c }{\cos} {\theta}_{{\superpartner{c}}} 
& -{\sin} {\theta}_{{\superpartner{c}}} \\
{\sin} {\theta}_{{\superpartner{c}}} 
& {\cos} {\theta}_{{\superpartner{c}}} \end{array} \right)
\left( {\myatop{{{\superpartner{c}}_{L}}}{{{\superpartner{c}}_{R}}}} \right),
\end{equation}
we obtain
\begin{eqnarray}
{\mathcal{M}} & = & {\frac{1}{i}} {\bra{{\xone} {\xone}}}
\left\{ {\xbarone} [ i |e| a_{u}
( {\cos} {\theta}_{{\superpartner{c}}} {\superpartner{c}}_{1}^{{\ast}}
+ {\sin} {\theta}_{{\superpartner{c}}} {\superpartner{c}}_{2}^{{\ast}} ) P_{L}
+ i |e| b_{u}
( {\cos} {\theta}_{{\superpartner{c}}} {\superpartner{c}}_{2}^{{\ast}}
- {\sin} {\theta}_{{\superpartner{c}}} {\superpartner{c}}_{1}^{{\ast}} ) P_{R}
] c \right\} \nonumber\\
 & & {\times} \left\{ {\bar{c}} [ -i |e| a_{u}
( {\cos} {\theta}_{{\superpartner{c}}} {\superpartner{c}}_{1}
+ {\sin} {\theta}_{{\superpartner{c}}} {\superpartner{c}}_{2} ) P_{R}
- i |e| b_{u}
( {\cos} {\theta}_{{\superpartner{c}}}  {\superpartner{c}}_{2}
- {\sin} {\theta}_{{\superpartner{c}}}  {\superpartner{c}}_{1} ) P_{L} ] 
\xone \right\} {\ket{{J/{\psi}}}} \nonumber\\
 & = & {\frac{{e^{2} F_{{J/{\psi}}} m_{{J/{\psi}}}}}{8}}
[ \bar{u}_\xone( k_1 ) {\slashed{{\epsilon}}}_{{J/{\psi}}}
{\gammabar} v_\xone( k_2 )
- \bar{u}_\xone(k_2) {\slashed{{\epsilon}}}_{{J/{\psi}}}
{\gammabar} v_\xone( k_{1} ) ] \nonumber\\
 & & {\times} \left(
{\frac{{( a_{u} {\cos} {\theta}_{{\superpartner{c}}} )^{2}
- ( b_{u}
{\sin} {\theta}_{{\superpartner{c}}} )^{2}}}{{m_{{\superpartner{c}}_{1}}^{2}}}}
+ {\frac{{( a_{u} {\sin} {\theta}_{{\superpartner{c}}}  )^{2}
- ( b_{u}
{\cos} {\theta}_{{\superpartner{c}}} )^{2}}}{{m_{{\superpartner{c}}_{2}}^{2}}}}
\right),
\end{eqnarray}
since ${\bra{0}} {\bar{c}} P_{{L/R}} c {\ket{{J/{\psi}}}} = 0$.  Hence
\begin{eqnarray}
{\overline{{| {\mathcal{M}} |^{2}}}} & = &
{\frac{{e^{4} F_{{J/{\psi}}}^{2}}}{{48}}} 
\left( {\frac{{( a_{u} {\cos} {\theta}_{{\superpartner{c}}} )^{2}
- ( b_{u}
{\sin} {\theta}_{{\superpartner{c}}} )^{2}}}{{m_{{\superpartner{c}}_{1}}^{2}}}}
+ {\frac{{( a_{u} {\sin} {\theta}_{{\superpartner{c}}} )^{2} 
- ( b_{u}
{\cos} {\theta}_{{\superpartner{c}}} )^{2}}}{{m_{{\superpartner{c}}_{2}}^{2}}}}
\right)^{2} \nonumber\\
 & & {\times} ( 2 {\dotproduct{{k_{1}}}{{p_{{J/{\psi}}}}}}
{\dotproduct{{k_{2}}}{{p_{{J/{\psi}}}}}}
+ m_{{J/{\psi}}}^{2} {\dotproduct{{k_{1}}}{{k_{2}}}}
- 3 m_{{J/{\psi}}}^{2} {\mxonesq} )
\end{eqnarray}
and so we obtain
\begin{eqnarray}
{\Gamma}_{{J/{\psi} {\to} {\xone} {\xone}}} & = &
{\frac{9}{{128}}} {\Gamma}_{{J/{\psi} {\to} e^{+} e^{-}}} m_{{J/{\psi}}}^{4}
\left[ 1 - 4
\left( {\frac{{\mxone}}{{m_{{J/{\psi}}}}}} \right)^{2} \right]^{{( 3 / 2 )}} 
\nonumber\\
 & & {\times} \left(
{\frac{{( a_{u} {\cos} {\theta}_{{\superpartner{c}}} )^{2}
- ( b_{u}
{\sin} {\theta}_{{\superpartner{c}}} )^{2}}}{{m_{{\superpartner{c}}_{1}}^{2}}}}
+ {\frac{{( a_{u} {\sin} {\theta}_{{\superpartner{c}}} )^{2}
- ( b_{u}
{\cos} {\theta}_{{\superpartner{c}}} )^{2}}}{{m_{{\superpartner{c}}_{2}}^{2}}}}
\right)^{2}.
\end{eqnarray}

Our results agrees with the decay width for orthotoponium decay to a
photino pair in Ref.~\cite{Ellis:1983ed}, with the replacements
$a_{u/d}, b_{u/d} {\to} e_{u} = 2 / 3,
{\theta}_{{\superpartner{{c}}}/{\superpartner{{b}}}} {\to} {\tilde{{\theta}}}_{t},
m_{{\superpartner{{c}}}/{\superpartner{{b}}}} {\to} m_{{\superpartner{{t}}}}$
for the case $m_{{\superpartner{{\gamma}}}} {\to} 0$, and assuming
$m_{{\theta}} = 2 m_{t}$.

The greatest branching ratio occurs when
${\theta}_{{\superpartner{c}}} = {\pi} / 2$ and the heavier charm
squark is decoupled.  In this limit, and using $m_{{J/{\psi}}} = 3.10$ GeV,
${\Gamma}_{J/{\psi} {\to} e^{+} e^{-}} / {\Gamma}_{J/{\psi}, \mathrm{tot}} =
5.94 {\times} 10^{-2}$~\cite{PDG}, the branching ratio for
$J/{\psi} {\to} {\xone} {\xone}$ is
\begin{equation}
\mbox{BR}( J/{\psi} {\to} {\xone} {\xone} ) = ( 5.12 {\times} 10^{-9} ) 
\left( {\frac{{100 \mbox{ GeV}}}{{m_{{{\superpartner{c}}_{1}}}}}} \right)
^{4} \left[ 1 - ( 4.17 {\times} 10^{-7} ) \left( {\frac{{\mxone}}{{1 
\mbox{ MeV}}}} \right)^{2} \right]^{{( 3 / 2 )}},
\end{equation}
which is maximized for ${\mxone} = 0$.  Setting
$m_{{{\superpartner{c}}_{1}}}$ to $100$ GeV, we find a branching ratio of
$5.12 {\times} 10^{-9}$.  This is well below the experimental upper
bound of $5.9 {\times} 10^{-4}$~\cite{Ablikim:2007ek}, but not very much
below the branching ratio for the SM process
$J/{\psi} {\to} {\nu} {\bar{{\nu}}}$ of
$2.70 {\times} 10^{-8}$~\cite{Chang:1997tq}.

Analogously,
\begin{eqnarray}
{\Gamma}_{{{\Upsilon} {\to} {\xone} {\xone}}} & = &
{\frac{9}{{32}}} {\Gamma}_{{{\Upsilon} {\to} e^{+} e^{-}}} m_{{\Upsilon}}^{4}
\left[ 1 - 4
\left( {\frac{{\mxone}}{{m_{{\Upsilon}}}}} \right)^{2} \right]^{{( 3 / 2 )}}
\nonumber\\
 & & {\times} \left(
{\frac{{( a_{d} {\cos} {\theta}_{{\superpartner{b}}} )^{2}
- ( b_{d}
{\sin} {\theta}_{{\superpartner{b}}} )^{2}}}{{m_{{\superpartner{b}}_{1}}^{2}}}}
+ {\frac{{( a_{d} {\sin} {\theta}_{{\superpartner{b}}} )^{2}
- ( b_{d}
{\cos} {\theta}_{{\superpartner{b}}} )^{2}}}{{m_{{\superpartner{b}}_{2}}^{2}}}}
\right)^{2}\,.
\end{eqnarray}
Here ${\theta}_{{\superpartner{b}}}$ is the mixing angle in the bottom
squark sector.  The greatest branching ratio occurs when
${\theta}_{{\superpartner{b}}} = {\pi} / 2$ and the heavier bottom
squark is infinitely heavy.  In this limit, the branching ratio for
${\Upsilon} {\to} {\xone} {\xone}$ is
\begin{equation}
\mbox{BR}( \Upsilon \to \xone \xone ) = ( 4.47 \times 10^{-8} )
\left( \frac{100 \mbox{ GeV}}{m_{\tilde b_1}} \right)^{4}
\left[ 1 - ( 4.69 {\times} 10^{-8} )
\left( {\frac{{\mxone}}{{1 \mbox{ MeV}}}} \right)^{2} \right]^{{( 3 / 2 )}},
\end{equation}
which is maximized for ${\mxone} = 0$. Here we have used
$m_{{\Upsilon}} = 9.46$ GeV,
${\Gamma}_{{\Upsilon} {\to} e^{+} e^{-}} / {\Gamma}_{{\Upsilon}, \mathrm{tot}} =
2.38 {\times} 10^{-2}$~\cite{PDG}.  Setting the lighter bottom squark
mass $m_{{\superpartner{b}}_{1}}$ to $100$ GeV, we find a branching ratio of
$4.47 {\times} 10^{-8}$, which is well below the experimental upper
bound of $2.5 {\times} 10^{-3}$~\cite{Tajima:2006nc}, and also well below
the branching ratio for the SM process
${\Upsilon} {\to} {\nu} {\bar{{\nu}}}$ of
$1.05 {\times} 10^{-5}$~\cite{Chang:1997tq}.

We note that ${\Upsilon} {\to}$ invisible has been used to probe light
dark matter through inversion of the annihilation
cross-section~\cite{McElrath:2005bp}.

\section{${\bm{\Ktopi}}$ and ${\bm{\BtoK}}$ within the minimal flavor violating MSSM}
\label{MFV_section}

As mentioned in Sect.~\ref{sec:generic_P_decay_section}, the two-body
decays of pseudoscalar mesons all have widths proportional to the
square of the mass of the $\xone$, and so there are no bounds for the
case of massless neutralinos.  This is not the case for three-body final
states, and so we examine whether there are significant bounds from
the decays ${\Ktopi}$ and ${\BtoK}$, where the SM analogues,
$K^{-} {\to} {\pi}^{-} {\nu} {\bar{{\nu}}}$ and
$B^{-} {\to} K^{-} {\nu} {\bar{{\nu}}}$, have very low branching
ratios~\cite{SM_prediction,Anisimovsky:2004hr, Adler:2008zza, BELLE:2007zk}. 

We present the results for the case of massless neutralinos.  The phase
space is smaller as the ${\xone}$ mass increases, and any terms in the
matrix-element-squared proportional to the ${\xone}$ mass are
subdominant.

\subsection{General analysis}

\begin{figure}
\begin{center}
\includegraphics[width=4cm]{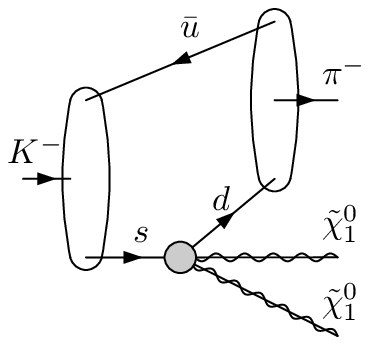}
\caption{Generic Feynman diagram of the decay ${\Ktopi}$.}
\label{K_to_pi_blobs}
\end{center}
\end{figure}

\begin{figure}
\begin{center}
\includegraphics[width=12cm]{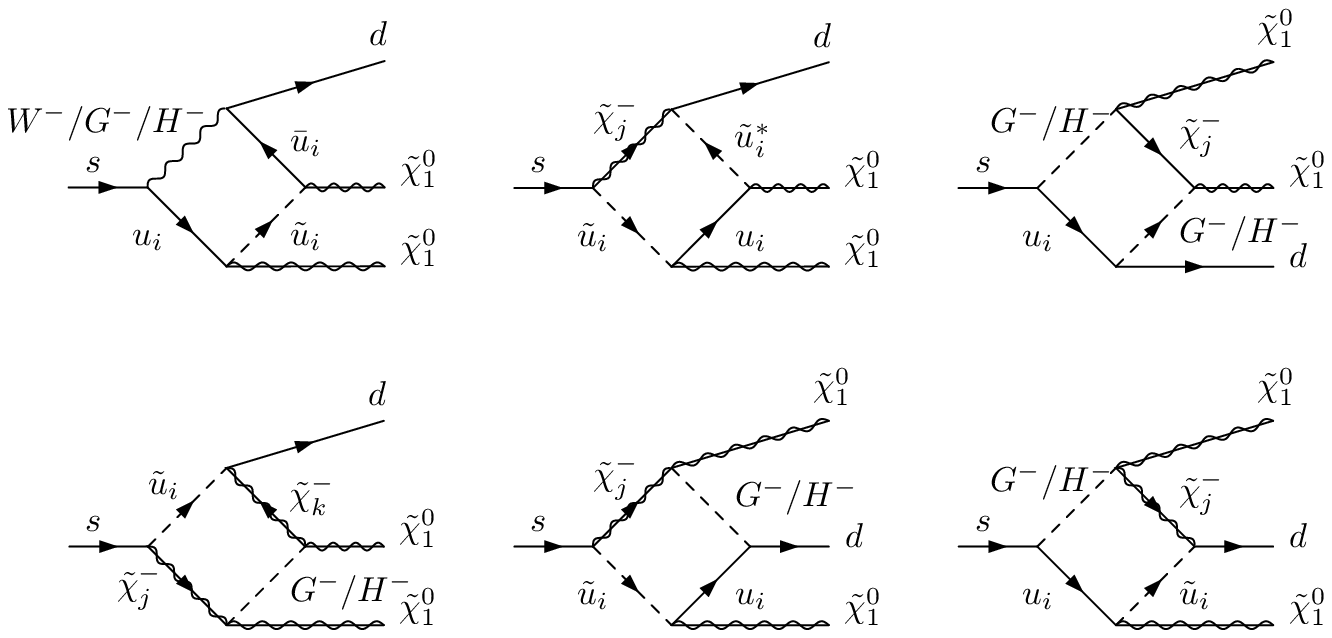}
\caption{Box Feynman diagrams for the decay $s {\to} d {\xone} {\xone}$.}
\label{s_to_d_binos_boxes}
\end{center}
\end{figure}

\begin{figure}
\begin{center}
\includegraphics[width=12cm]{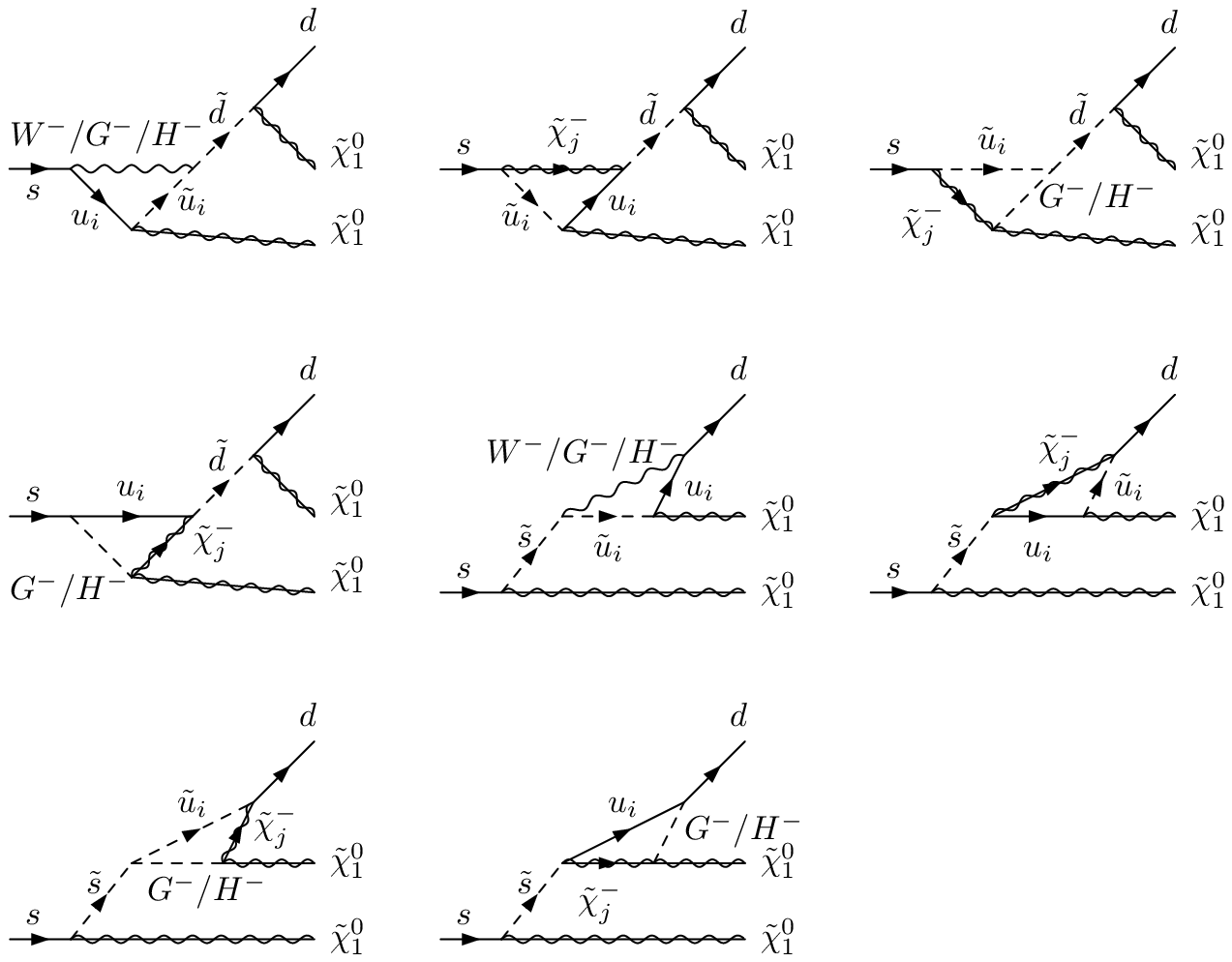}
\caption{Triangle Feynman diagrams for the decay $s {\to} d {\xone} {\xone}$.}
\label{s_to_d_binos_triangles}
\end{center}
\end{figure}

\begin{figure}
\begin{center}
\includegraphics[width=12cm]{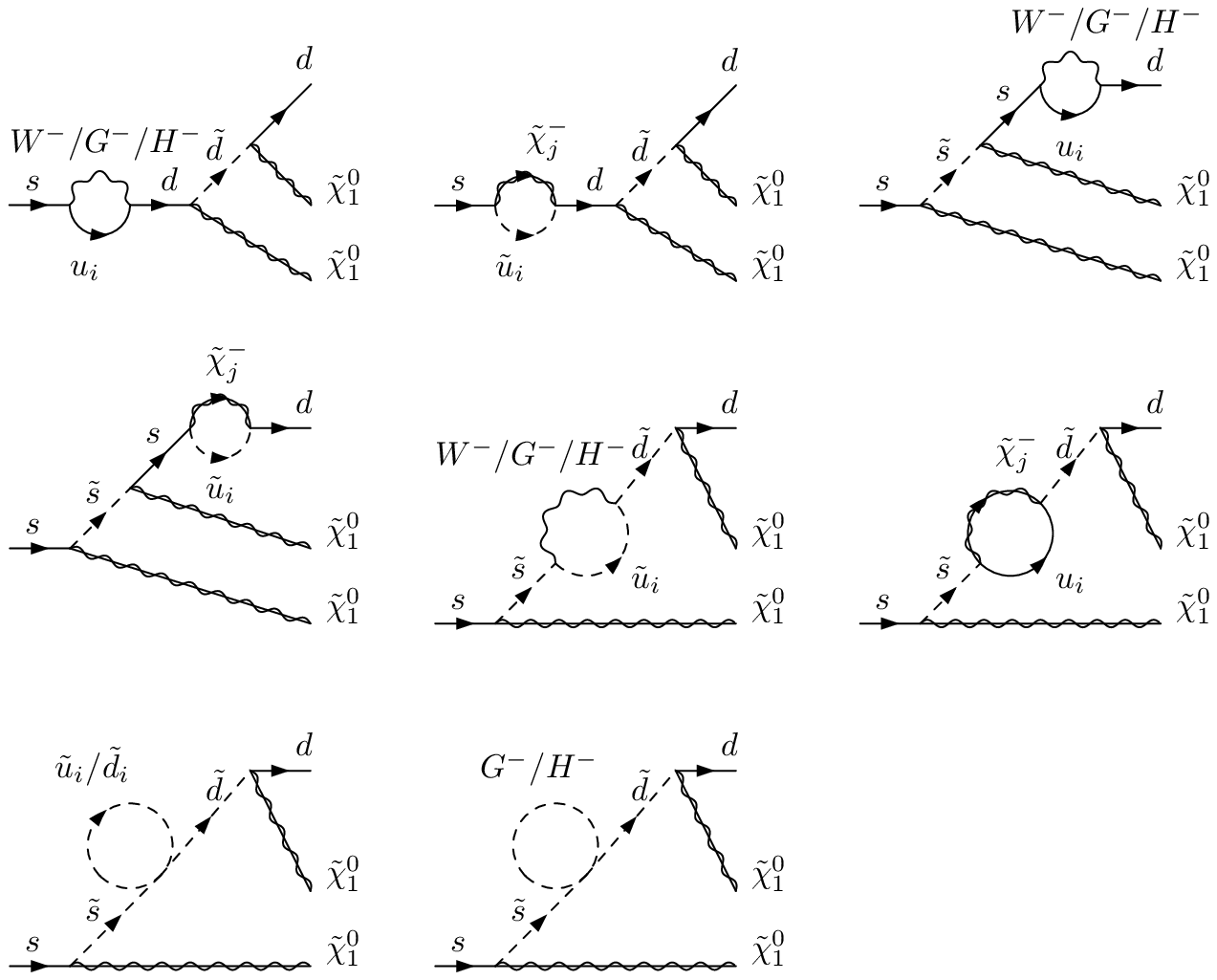}
\caption{Circle Feynman diagrams for the decay $s {\to} d {\xone} {\xone}$.}
\label{s_to_d_binos_circles}
\end{center}
\end{figure}

For this section, we impose minimal flavor violation 
(MFV)~\cite{Gabrielli:1994ff} at the low energy scale at which we 
are working (\ie\ at the mass of the decaying meson).  MFV models 
are defined by the requirement that all flavor violating and 
CP-violating transitions are described by the
Cabibbo-Kobayashi-Maskawa (CKM) matrix of the SM, 
see \eg~Refs.~\cite{Buras:2003jf,Jager:2008fc} and references therein.  
Thus the decays ${\Ktopi}$ and ${\BtoK}$ only have leading 
contributions from one-loop diagrams. Those for
${\Ktopi}$ are shown in Figs.~\ref{K_to_pi_blobs} to
\ref{s_to_d_binos_circles}, without the diagrams with the neutralinos
crossed. Through the loops, the partial width depends on many
parameters, and not in a simple way.  Therefore we have opted to use a
set of supersymmetric benchmark scenarios, thereby fixing the masses
and couplings.  We use the masses and mixings of the Snowmass points
SPS1a, SPS2, SPS3, SPS4 and SPS5~\cite{SPS}, calculated using
\SOFTSUSY~\cite{SOFTSUSY}. We have modified the neutralino sector of 
these benchmark points by choosing $M_{1}$ according to 
Eq.~(\ref{eq:m1_min}) such that the lightest neutralino is bino-like 
and massless.  The resulting supersymmetric models we denote
`pseudo-SPS' points.  The imposition of MFV and the degeneracy of the
up- and charm-squark masses leads to a super-Glashow-Iliopoulos-Maiani
(super-GIM)~\cite{superGIM} suppression in the ${\Ktopi}$ decay and to
a lesser extent in the ${\BtoK}$ decay.

Since the scale of the sparticle masses and the $W$ boson mass are so
much greater than the masses of the kaon or $B$ meson, we integrate
out the heavy degrees of freedom to obtain effective four-fermion
vertices~\cite{Buras:1998raa}.  We then write the amplitude for
${\Ktopi}$ as
\begin{eqnarray}\label{eq:ktopi}
{\mathcal{M}}^{{\text{MFV}}}_{{\Ktopi}} & = & {\frac{{G_{F}}}{2}} \left( C_{{\Slashed{F}}} {\bar{u}}( k_{1} ) {\gamma}_{{\mu}} {\gammabar} v( k_{2} ) {\bra{{{\pi}^{-}}}} {\bar{d}} {\gamma}^{{\mu}} s {\ket{{K^{-}}}} + {\frac{{C_{{\slashed{p}}}}}{{m_{K}}}} {\bar{u}}( k_{1} ) {\slashed{p}}_{s} v( k_{2} ) {\bra{{{\pi}^{-}}}} {\bar{d}} s {\ket{{K^{-}}}} \right. \nonumber\\
& & \left. + C_{L} {\bar{u}}( k_{1} ) P_{L} v( k_{2} ) {\bra{{{\pi}^{-}}}} {\bar{d}} s {\ket{{K^{-}}}} + C_{R} {\bar{u}}( k_{1} ) P_{R} v( k_{2} ) {\bra{{{\pi}^{-}}}} {\bar{d}} s {\ket{{K^{-}}}} \vphantom{\frac{G}{2}} \right) - ( k_{1} {\leftrightarrow} k_{2} ), \nonumber\\
& & 
\end{eqnarray}
recalling that $k_{1}$ and $k_{2}$ are the momenta of the
(indistinguishable) final-state neutralinos.  The $C_i$ coefficients,
$i = {\Slashed{F}},{\slashed{p}}, L, R$, are dimensionless.  These
are the only terms possible, since
${\xbarone} {\gamma}^{{\mu}} {\xone}$ and
${\xbarone} [ {\gamma}^{{\mu}}, {\gamma}^{{\nu}} ] {\xone}$ are
identically zero for Majorana fermions.
Conservation of momentum and the Dirac equation for the spinors
can be used to reduce any other terms to the set above, [applying the
identities
${\bar{u}}( k_{2} ) {\gamma}^{{\mu}} {\gammabar} v( k_{1} ) =
v^{T}( k_{2} ) {\gammabar} {\gamma}^{T{\mu}} {\bar{u}}^{T}( k_{2} ) =
-{\bar{u}}( k_{1} ) {\gamma}^{{\mu}} {\gammabar} v( k_{2} )$
and ${\bar{u}}( k_{2} ) P_{L/R} v( k_{1} )
= v^{T}( k_{2} ) P_{L/R} {\bar{u}}^{T}( k_{2} )
= -{\bar{u}}( k_{1} ) P_{L/R} v( k_{2} )$].

The coefficients $C_{i}$ were calculated with the aid of
\FormCalc~\cite{FormCalc}, using techniques detailed in
Ref.~\cite{FormCalc_tricks}, with the approximation of neglecting all
external momenta compared to the sparticle and $W$ boson masses.  The
loop integrals in this approximation are straightforward and can be
written as powers and logarithms of masses, so numerical
instabilities in the reduction of tensor integrals with extreme scale
differences are avoided.  Variations of these coefficients over the
phase space of the decay are of order $m_{K}^{2} / m_{W}^{2}$ and
therefore we ignore them.  This is analogous to the approach taken
in Ref.~\cite{Inami:1980fz} in calculating the leading order value
for the standard model decay $K^{+} {\to} {\pi}^{0} e^{+} {\nu}$.

Our calculation takes into account all diagrams at the one-loop level
and is not restricted by requiring the squark masses to be small
compared to the mass of the $W$ boson, thus improving the
original calculations for photino-like ${\xone}$ of
Refs.~\cite{Gaillard:1982rw,Ellis:1982ve}.

The relevant terms for ${\BtoK}$ can be obtained from the above by 
replacing the mesons in the bra and ket appropriately and the $s$
quark with a $b$ quark, and the ${\bar{d}}$ with a ${\bar{s}}$.

We use the mesonic form factors as given in Ref.~\cite{Nam:2007fx}
for ${\Ktopi}$ and Ref.~\cite{Ball:2004hn,Ball:2004ye} 
for  ${\BtoK}$ (reproduced for convenience in
App.~\ref{form_factors}),
but for the moment, we shall use the shorthand for the 
mesonic current ${\bra{{{\pi}^{-}}}} {\bar{d}} {\gamma}^{{\mu}} s 
{\ket{{K^{-}}}} = F^{{\mu}}$.  Taking the dot product of both sides 
with the difference of the momenta of the kaon and the pion, 
which to a good approximation is equal to the difference in momenta 
of the strange quark and the down quark, and using the Dirac equation 
for the quarks allows us to write 
\begin{equation}
{\bra{{\pi^-}}} \bar{d} s {\ket{{K^{-}}}} =
{\frac{{\dotproduct{F}{{( p_K - p_\pi )}}}}{( m_{s} + m_{d} )}}.
\end{equation}
In this notation
\begin{eqnarray}
{\mathcal{M}}^{{\text{MFV}}}_{{\Ktopi}} & = & G_{F} \left( C_{{\Slashed{F}}} {\bar{u}}( k_{1} ) {\Slashed{F}} {\gammabar} v( k_{2} ) + C_{{\slashed{p}}} {\bar{u}}( k_{1} ) {\slashed{p}}_{s} {\gammabar} v( k_{2} ) {\frac{{{\dotproduct{F}{{( p_{K} - p_{{\pi}} )}}}}}{{m_{K} ( m_{s} + m_{d} )}}} \right. \nonumber\\
  & & \left. + C_{L} {\bar{u}}( k_{1} ) P_{L} v( k_{2} ) {\frac{{{\dotproduct{F}{{( p_{K} - p_{{\pi}} )}}}}}{{( m_{s} + m_{d} )}}} + C_{R} {\bar{u}}( k_{1} ) P_{R} v( k_{2} ) {\frac{{{\dotproduct{F}{{( p_{K} - p_{{\pi}} )}}}}}{{( m_{s} + m_{d} )}}} \right).
\end{eqnarray}

As can be seen in Sect.~\ref{MFV_numerical_results_section}, the
coefficient $C_{{\Slashed{F}}}$ is much larger than the other coefficients for
the benchmark points we use for the decay ${\Ktopi}$.  If we neglect all the
$C$ coefficients except $C_{{\Slashed{F}}}$, we find that the spin-averaged
matrix-element-squared for ${\Ktopi}$ is
\begin{equation}
{\overline{{|{\mathcal{M}}^{\text{MFV}}_{\Ktopi}|^{2}}}}
 = 4 G_{F}^{2} |C_{{\Slashed{F}}}|^{2} \left(
{\dotproduct{{F^{{\ast}}}}{{k_{1}}}} {\dotproduct{F}{{k_{2}}}} + 
{\dotproduct{{F^{{\ast}}}}{{k_{2}}}} {\dotproduct{F}{{k_{1}}}} - 
{\dotproduct{{F^{{\ast}}}}{F}} {\dotproduct{{k_{1}}}{{k_{2}}}} \right),
\end{equation}
which differs from that for $K^{-} {\to} {\pi}^{0} e {\bar{{\nu}}}$
(which has a branching ratio of $5.08 {\times} 10^{-2}$~\cite{PDG})
only by an overall factor [assuming a massless electron and isospin
symmetry for the up and down valence quarks; the factor of
$( 1 / {\sqrt{2}} )^{2}$ from the coefficient of $u {\bar{u}}$ in
${\pi}^{0}$ balances the factor of $1 / 2$ from identical final-state
particles in the ${\Ktopi}$ case].  In this case the behavior over
the phase space is identical and the ratio of the decay widths is
equal to the ratio of $| C_{{\Slashed{F}}} |^{2}$ to $| V_{us} |^{2}$.
For the case of ${\BtoK}$, we see that $C_{{\Slashed{F}}}$ is not as
dominant as for ${\Ktopi}$, and we present the results from directly
evaluating the decay width using the expressions for $F^{{\mu}}$ given
in Ref.~\cite{Ball:2004hn,Ball:2004ye}\ (though with diminished
accuracy from using mesonic form factors).  Direct evaluation was
found to agree at the percent level for ${\Ktopi}$.  For
${\BtoK}$, $C_{R}$ is comparable to $C_{{\Slashed{F}}}$ and the effect
of neglecting it decreases the branching ratio from by roughly
$0.3\%$ for pseudo-SPS5 up to by $66\%$ for pseudo-SPS4.

\subsection{Numerical results}
\label{MFV_numerical_results_section}

We present the numerical values of the $C$ coefficients in
Tables~\ref{Ktopi_numerical_table}\ and \ref{BtoK_numerical_table}.
In the final column of each table, headed BR/exp, we present the ratio
of either the computed ${\Ktopi}$ branching ratio to the experimental
value of the branching ratio for
$K^{-} {\to} {\pi}^{-} {\nu} {\bar{{\nu}}}$ (which is
$1.73 {\times} 10^{-10}$~\cite{Artamonov:2009sz}), or
the computed branching ratio of ${\BtoK}$ to the current experimental
upper bound on the branching ratio for
$B^{-} {\to} K^{-} {\nu} {\bar{{\nu}}}$
(which is $1.4 {\times} 10^{-5}$~\cite{BELLE:2007zk}).

\begin{center}
\begin{table}[ht]
\renewcommand{\arraystretch}{1.5}
\begin{tabular}{l d@{${\columnspacefixer}$}l d@{${\columnspacefixer}$}l d@{${\columnspacefixer}$}l d@{${\columnspacefixer}$}l d@{${\columnspacefixer}$}l d@{${\columnspacefixer}$}l}
\hline \\[-10mm] \hline
pseudo- & & & & & & & & & & & & \\[-6mm]
SPS & \multicolumn{2}{c}{$|C_{{\Slashed{F}}}|$} & \multicolumn{2}{c}{$|C_{{\slashed{p}}}|$} & \multicolumn{2}{c}{$|C_{L}|$} & \multicolumn{2}{c}{$|C_{R}|$} & \multicolumn{2}{c}{BR} & \multicolumn{2}{c}{BR/exp}
\newhline
1a  & $5.70$ & ${\times} 10^{-8}$ & $7.45$ & ${\times} 10^{-14}$ & $4.93$ & ${\times} 10^{-12}$ & $2.60$ & ${\times} 10^{-10}$ & $3.28$ & ${\times} 10^{-16}$ & $1.90$ & ${\times} 10^{-6}$
\\
2   & $3.81$ & ${\times} 10^{-9}$ & $7.77$ & ${\times} 10^{-14}$ & $9.55$ & ${\times} 10^{-12}$ & $4.10$ & ${\times} 10^{-11}$ & $1.47$ & ${\times} 10^{-18}$ & $8.49$ & ${\times} 10^{-9}$
\\
3   & $2.63$ & ${\times} 10^{-8}$ & $3.74$ & ${\times} 10^{-14}$ & $5.13$ & ${\times} 10^{-12}$ & $1.25$ & ${\times} 10^{-10}$ & $6.99$ & ${\times} 10^{-17}$ & $4.04$ & ${\times} 10^{-7}$
\\
4   & $2.95$ & ${\times} 10^{-8}$ & $6.41$ & ${\times} 10^{-14}$ & $1.34$ & ${\times} 10^{-12}$ & $1.42$ & ${\times} 10^{-9}$ & $8.76$ & ${\times} 10^{-17}$ & $5.06$ & ${\times} 10^{-7}$
\\
5   & $7.12$ & ${\times} 10^{-8}$ & $2.81$ & ${\times} 10^{-14}$ & $5.61$ & ${\times} 10^{-12}$ & $7.11$ & ${\times} 10^{-11}$ & $5.12$ & ${\times} 10^{-16}$ & $2.96$ & ${\times} 10^{-6}$
\newhline \\[-10mm] \hline
\end{tabular}
\caption{Numerical values for the coefficients $C_{i}$ defined in
  Eq.~(\ref{eq:ktopi}) for ${\Ktopi}$ at the various pseudo-SPS points
  described in the text. The final column shows the ratio of the
  branching ratio for ${\Ktopi}$ to the experimental value of the
  branching ratio for $K^{-} {\to} {\pi}^{-} {\nu} {\bar{{\nu}}}$
  ($1.73 {\times} 10^{-10}$).}
\label{Ktopi_numerical_table}
\end{table}
\end{center}

\begin{center}
\begin{table}[ht]
\renewcommand{\arraystretch}{1.5}
\begin{tabular}{l d@{${\columnspacefixer}$}l d@{${\columnspacefixer}$}l d@{${\columnspacefixer}$}l d@{${\columnspacefixer}$}l d@{${\columnspacefixer}$}l d@{${\columnspacefixer}$}l}
\hline \\[-10mm] \hline
pseudo- & & & & & & & & & & & & \\[-6mm]
SPS & \multicolumn{2}{c}{$|C_{{\Slashed{F}}}|$} & \multicolumn{2}{c}{$|C_{{\slashed{p}}}|$} & \multicolumn{2}{c}{$|C_{L}|$} & \multicolumn{2}{c}{$|C_{R}|$} & \multicolumn{2}{c}{BR} & \multicolumn{2}{c}{BR/exp}
\newhline
1a  & $6.49$ & ${\times} 10^{-6}$ & $3.77$ & ${\times} 10^{-9}$ & $1.93$ & ${\times} 10^{-8}$ & $7.22$ & ${\times} 10^{-7}$ & $3.35$ & ${\times} 10^{-10}$ & $2.39$ & ${\times} 10^{-5}$
\\
2   & $5.44$ & ${\times} 10^{-7}$ & $3.69$ & ${\times} 10^{-9}$ & $4.13$ & ${\times} 10^{-8}$ & $1.83$ & ${\times} 10^{-7}$ & $2.48$ & ${\times} 10^{-12}$ & $1.77$ & ${\times} 10^{-7}$
\\
3   & $3.00$ & ${\times} 10^{-6}$ & $1.87$ & ${\times} 10^{-9}$ & $2.16$ & ${\times} 10^{-8}$ & $4.54$ & ${\times} 10^{-7}$ & $7.19$ & ${\times} 10^{-11}$ & $5.14$ & ${\times} 10^{-6}$
\\
4   & $3.31$ & ${\times} 10^{-6}$ & $3.22$ & ${\times} 10^{-9}$ & $5.85$ & ${\times} 10^{-9}$ & $6.38$ & ${\times} 10^{-6}$ & $2.53$ & ${\times} 10^{-10}$ & $1.81$ & ${\times} 10^{-5}$
\\
5   & $9.50$ & ${\times} 10^{-6}$ & $1.49$ & ${\times} 10^{-9}$ & $2.12$ & ${\times} 10^{-8}$ & $2.27$ & ${\times} 10^{-7}$ & $7.14$ & ${\times} 10^{-10}$ & $5.10$ & ${\times} 10^{-5}$
\newhline \\[-10mm] \hline
\end{tabular}
\caption{Numerical values for the coefficients $C_{i}$ defined in Eq.~(\ref{eq:ktopi}) for ${\BtoK}$ at the various pseudo-SPS points described in the text.  The final column shows the ratio of the branching ratio for ${\BtoK}$ to the current experimental upper bound on the branching ratio for $B^{-} {\to} K^{-} {\nu} {\bar{{\nu}}}$ ($1.4 {\times} 10^{-5}$).} 
\label{BtoK_numerical_table}
\end{table}
\end{center}

We see that the branching ratios for the decays ${\Ktopi}$ and
${\BtoK}$ in the MSSM with MFV are several orders of magnitude smaller
than the respective standard model processes, with no chance to
observe them in the experiments of today and the
foreseeable future.

The strong relative suppression of the kaon and $B$ meson decays to
neutralinos compared to their respective SM decays to neutrinos can be
explained as a combination of various factors.  First, the sparticle
masses of the SPS benchmark scenarios are considerably larger than the
$W$ mass; in particular, the SPS2 benchmark point with squark masses
of ${\cal O}(1.5)$~TeV is most strongly suppressed.  Moreover, the SM
process has four $SU(2)$ vertices while the MSSM process has two
$SU(2)$ vertices and two $U(1)$ vertices, resulting in a further
suppression of the MSSM decay matrix element proportional to
${\tan}^{2} {\theta}_{W}$.  Since the flavor-changing aspect is carried
by the $W$ boson or the charged Higgs bosons, the quarks/squarks are
mainly left-handed, and their couplings to the bino-like neutralinos
are suppressed by their small hypercharge.  Finally, the spin structure
of the MSSM decay amplitude provides another order-of-magnitude
suppression with respect to the SM decay.  Considering all these
factors together leads to a suppression of several orders of magnitude.

\section{$\bm{K^- \rightarrow \pi^- \tilde{\chi}_1^0
    \tilde{\chi}_1^0}$ and $\bm{B^- \rightarrow K^-/\pi^-
    \tilde{\chi}_1^0 \tilde{\chi}_1^0}$ within the non-minimal flavor
  violating MSSM}
\label{sec:non-mfv}

Going beyond the previous section, we here consider non-minimal flavor
violation in the MSSM, for which there are many potential sources. For
example, the squark mass matrix in the soft-breaking part of the
Lagrangian in general couples squarks of different flavor to each
other, resulting in flavor changing neutral currents (FCNCs).  Thus
from the theoretical point of view there is \textit{a priori} no
reason to assume MFV within the MSSM.

To deal with the complicated flavor structure of the MSSM, we employ
the mass insertion approximation \cite{Hall:1985dx,Jager:2008fc},
which is defined in the super-CKM basis
\cite{Jager:2008fc,Dugan:1984qf,Hall:1985dx}. In this basis, the
Lagrangian is written in the quark mass eigenstate basis. The squarks
are rotated with the same rotation matrices as the quarks. This leads
to a diagonal flavor structure of the gauge interactions, but the
resulting squarks are not necessarily mass eigenstates. The squark
propagator can be expanded according to Ref.~\cite{Chankowski:2005jh}:
\begin{eqnarray}
\left< \tilde{q}_\alpha \tilde{q}_\beta^* \right> &=& \frac{i}{(k^2 
\mathds{1} - \tilde{m}^2 \mathds{1} - \delta m^2)_
{\alpha \beta}}  \nonumber \\
&=& \frac{i}{k^2-\tilde{m}^2} \mathds{1}_{\alpha \beta} 
+ \frac{i}{(k^2-\tilde{m}^2)^2} \delta m^2_{\alpha \beta} 
+ \mathcal{O}(\delta m^4) ,
\label{mass_expansion}
\end{eqnarray}
where $\mathds{1}$ is the identity matrix.  $\alpha$ and $\beta$ are
the indices of the $6 \times 6$ squark mass-squared matrix
$\,\tilde{M}^2$ and
$\tilde{m}^2 =
\sqrt{(\tilde{M}^2)_{\alpha\alpha}(\tilde{M}^2)_{\beta \beta}}\,$
is the ``average'' squark mass squared.  The off-diagonal elements
of the squared squark mass matrix in the super-CKM basis are given
by $\delta m^2_{\alpha\beta}$.

To avoid large FCNCs, which would be in contradiction with
experimental observations, we shall make the common assumption
$\tilde{m}^2 \gg \delta m^2_{\alpha \beta}$~\cite{Jager:2008fc}. In
this case higher order terms in $\delta m^2_{\alpha \beta}$ in
Eq.~(\ref{mass_expansion}), denoted by $\mathcal{O}(\delta m^4)$, can
usually be neglected.  However, if the linear term in
$\delta m^2_{\alpha \beta}$ in Eq.~(\ref{mass_expansion}) vanishes,
higher orders in $\delta m^2$ need to be taken into account.

In what follows we parametrize the amount of flavor violation with the
help of the dimensionless mass insertion parameters~\cite{Jager:2008fc}:
\begin{equation}
(\delta^d_{ij})_{XY} \equiv \frac{(\delta m^2_{d\,XY})_{ij}}{\tilde{m}^2} 
\, , \qquad i \not = j \, ,
\label{mi_params}
\end{equation}
where $(\delta m^2_{d\,XY})_{ij}$ with $XY=LL, RR, LR, RL$ are
the off-diagonal elements of the $6 \times 6$ down-squark mass-squared
matrices, which couple squarks of helicity $X$ to squarks of helicity
$Y$.

{}From the non-observation of FCNCs beyond the standard
model contributions, bounds on the mass insertion parameters,
Eq.~(\ref{mi_params}), can be derived
\cite{Gabbiani:1996hi,Ciuchini:1998ix,Misiak:1997ei,Becirevic:2001jj,
  Foster:2006ze,Ball:2006xx,Ciuchini:2006dx,
  Borzumati:1999qt,Ciuchini:2002uv,Silvestrini:2007yf,Jager:2008fc}:
\begin{eqnarray}
|(\delta^d_{ds})_{LL/RR}| \lsim  \mathcal{O}(10^{-2})\,,\quad & 
|(\delta^d_{ds})_{RL/LR}| \lsim  \mathcal{O}(10^{-3})\,, \nonumber \\
|(\delta^d_{db})_{LL/RR}| \lsim  \mathcal{O}(10^{-1})\,,\quad & 
|(\delta^d_{db})_{RL/LR}| \lsim  \mathcal{O}(10^{-2})\,, \nonumber \\
|(\delta^d_{sb})_{LL/RR}| \lsim  \mathcal{O}(10^{-1})\,,\quad &
|(\delta^d_{sb})_{RL/LR}| \lsim  \mathcal{O}(10^{-2})\, .
\label{general_bounds}
\end{eqnarray}

Here an average squark mass of $\tilde{m}=500$ GeV is assumed.

The bounds in Eq.~(\ref{general_bounds}) were obtained mainly from
meson-antimeson mixing. In this case, they scale (at leading order)
with $\tilde{m}/(500 \, {\rm GeV})$ as long as the ratio of the squark
and gluino masses is fixed, \textit{i.e.} the bounds get weaker for
increased squark mass. Note that in addition to experimental
constraints, there exist theoretical constraints on all
$(\delta^d_{ij})_{LR}$ from the requirement of vacuum stability
\cite{Casas:1996de}. These bounds are comparable or even stronger than
those in Eq.~(\ref{general_bounds}).  Furthermore, they do not become
weaker as the SUSY scale is raised.

The constraints on the mass insertions which connect squarks
associated with different helicity are usually much more restrictive
than those on the ``helicity-conserving'' mass insertions.  We
therefore concentrate in the following on processes where flavor
violation is mediated by either only left-handed or only right-handed
squarks. We thus consider only contributions from either $\delta_{LL}$
or $\delta_{RR}$.

\subsection{Matrix elements and partial width}
\label{nonMFV_ME}

\begin{figure}
\begin{center}
\includegraphics[width=8cm]{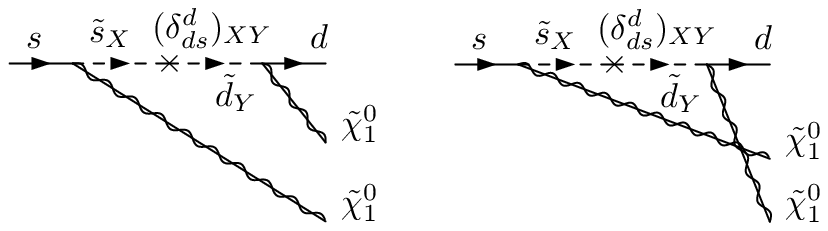}
\caption{\label{diag_nonMFV} The tree-level diagrams contributing to
  the process
  $K^- \rightarrow \pi^- \tilde{\chi}_1^0 \tilde{\chi}_1^0$ without
  assuming MFV.  The FCNCs are mediated by left- ($X, Y = L$) and
  right-handed ($X, Y = R$) squarks, respectively.} 
\end{center}
\end{figure}

We show in Fig.~\ref{diag_nonMFV} the tree-level Feynman diagrams
which allow for the process
$K^- \rightarrow \pi^- \tilde{\chi}_1^0 \tilde{\chi}_1^0$ assuming
a nearly massless bino-like neutralino $\tilde{\chi}_1^0$. We obtain,
after employing Fierz transformations, the following quark-level
matrix elements from the diagrams in Fig.~\ref{diag_nonMFV}:
\begin{eqnarray}
\mathcal{M}_{LL} &=& \frac{e^2 (\delta^d_{ds})_{LL}}{36 \cos^2\theta_W 
\tilde{m}^2} \;
[\bar d \gamma^\mu P_L s]\, [\bar u(k_1) \gamma_\mu \gamma_5 v(k_2)]
\, , \nonumber \\
\mathcal{M}_{RR} &=& \frac{- e^2 (\delta^d_{ds})_{RR}}{9 \cos^2 \theta_W 
\tilde{m}^2} \;
[\bar d \gamma^\mu P_R s] \,[\bar u(k_1) \gamma_\mu \gamma_5 v(k_2)] \, . 
\end{eqnarray}
Here $\mathcal{M}_{LL}$ and $\mathcal{M}_{RR}$ are the matrix
elements involving only left- and only right-handed squarks,
respectively. The quark currents are replaced by their corresponding
hadronic matrix elements, which are parametrized by the form factors
$f_+$ and $f_-$; see App.~\ref{form_factors} for details.

From the squared matrix elements, Eq.~(\ref{squared_ME}), we
easily obtain the partial width for
$K^- \rightarrow \pi^- \tilde{\chi}_1^0 \tilde{\chi}_1^0$.
Note that we have to introduce an additional factor of $1/2$
because this process involves two identical particles in the final state.

The decays
\begin{equation}
B^- \rightarrow \pi^-/ K^- \tilde{\chi}_1^0 \tilde{\chi}_1^0 \, , 
\label{nonMFV_Bdecays2}
\end{equation} 
are described by similar squared matrix elements, which can be
obtained from Eq.~(\ref{squared_ME}) by replacing the form factors
according to Eq.~(\ref{nonMFV_Bform_factors}) and,
if considering ${\BtoK}$, $p_\pi$ ($m_\pi$) by
$p_K$ ($m_K$).

\subsection{Branching ratios and excluded parameter space}

\begin{figure}
\begin{center}
\includegraphics[scale=1.0]{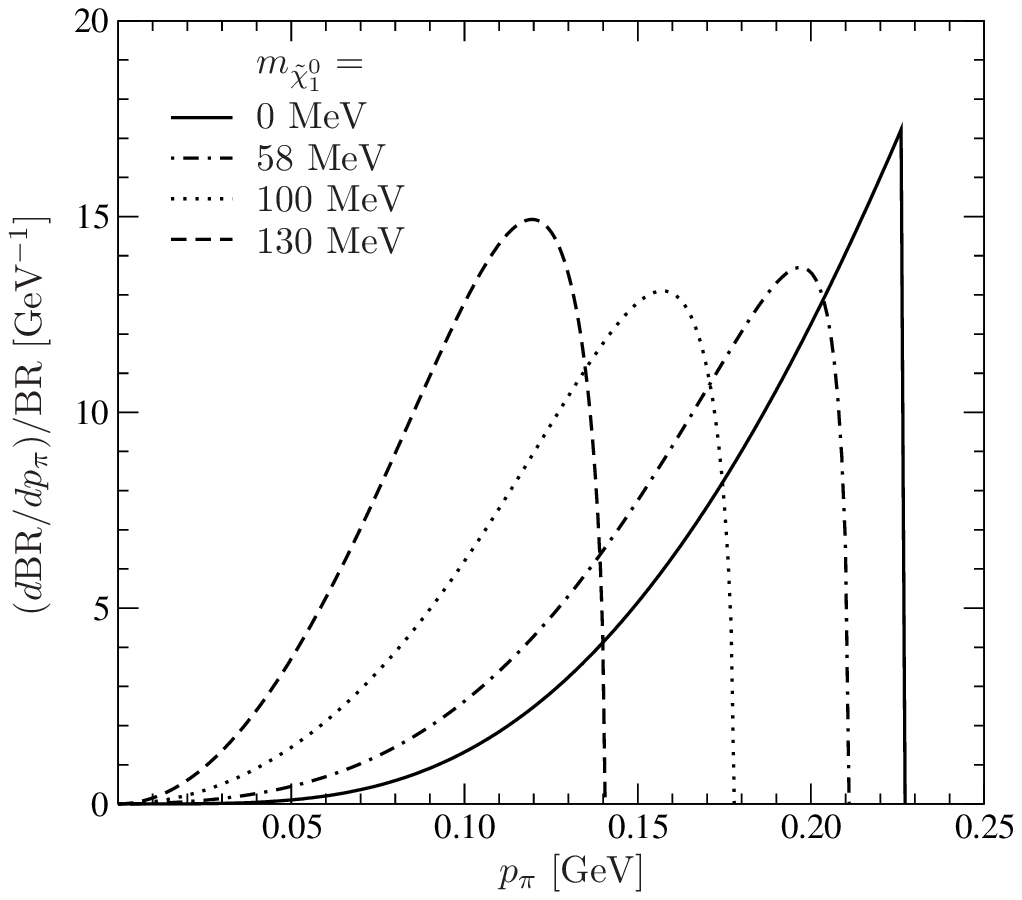}
\caption{\label{diff_BRs} Pion momentum distribution for the process
  $K^-\rightarrow \pi^- \tilde{\chi}_1^0 \tilde{\chi}_1^0$ for
  different neutralino masses $m_{\tilde{\chi}_1^0}$. The
  distributions are normalized to one.}
\end{center}
\end{figure}

The SM process $K^- \rightarrow \pi^- \nu \bar\nu$ has the same
experimental signature as the kaon decay into a pion plus neutralinos,
namely a charged pion and missing energy.  Here $\nu$ is a neutrino
of arbitrary flavor. The theoretical prediction for the branching
ratio for this decay (within the SM) is~\cite{SM_prediction}
\begin{equation}
\text{BR}(K^- \rightarrow \pi^- \nu \bar \nu)|_{\rm theory} = (8.5 \pm 0.7) \times 10^{-11}. 
\label{SM_predict}
\end{equation}

The squared matrix elements of the SM decay depend on the external
momenta in exactly the same way as those of the SUSY decay in
Eq.~(\ref{squared_ME}) for massless $\tilde{\chi}_1^0$s. The
distribution of the pion momentum $p_{\pi}$ for
$K^-\rightarrow \pi^- \tilde{\chi}_1^0 \tilde{\chi}_1^0\,$ is then
equal to the $p_{\pi}$-distribution of the SM process.  However,
this is no longer the case for massive $\tilde{\chi}_1^0$s as can
be seen in Fig.~\ref{diff_BRs}.

We show in Fig.~\ref{diff_BRs} the $p_{\pi}$-distribution
$(d\text{BR}/dp_\pi)/\text{BR}$ where
$\text{BR}=\text{BR}(K^-\rightarrow \pi^- \tilde{\chi}_1^0
\tilde{\chi}_1^0)$, for different values of $m_{\tilde{\chi}_1^0}$ in
the kaon rest frame.  The distributions are normalized to one.  We
have employed the squared matrix element of Eq.~(\ref{squared_ME2}) to
calculate $(d\text{BR}/dp_\pi)/\text{BR}$.  We see that for
$m_{\tilde{\chi}_1^0} > 130 $ MeV, $p_{\pi}$ is smaller than $140$
MeV. This has important consequences for experimental searches.

The E787 and E949 collaborations have observed events consistent with
the SM decay $K^- \rightarrow \pi^- \nu \bar \nu$. They found that
\cite{Artamonov:2009sz}
\begin{equation}
\text{BR}(K^- \rightarrow \pi^- \nu \bar \nu)|_{\rm exp.}
= (1.73^{+1.15}_{-1.05}) \times 10^{-10}, 
\label{SM_exp}
\end{equation}
assuming a $p_{\pi}$-spectrum equal to the SM prediction. To separate
the signal from the background, $p_{\pi}$ regions were selected,
namely $211$ MeV $< p_{\pi} < 229$ MeV (region I)
\cite{Adler:2008zza,Adler:2001xv} and $140$ MeV $< p_{\pi} < 199$ MeV
(region II) \cite{Adler:2004hp,Artamonov:2009sz}.  These were chosen
such that the background $K^- \rightarrow \pi^- \pi^0$ with $p_\pi
\approx 205$ MeV is excluded.  It follows from Fig.~\ref{diff_BRs}
that the experimental searches in region I (region II) were
insensitive to the process $K^-\rightarrow \pi^- \tilde{\chi}_1^0
\tilde{\chi}_1^0$ if $m_{\tilde{\chi}_1^0} > 58$ MeV
($m_{\tilde{\chi}_1^0} > 130$ MeV), because the respective pion
momenta are then too small.

In the following, we will estimate the experimental sensitivity for
scenarios with $m_{\tilde{\chi}_1^0} \not= 0$ with the help of the
correction factor
\begin{equation}
f_c \equiv \frac{I(m_{\tilde{\chi}_1^0} \not= 0 )}
{I(m_{\tilde{\chi}_1^0}=0 )}
\label{corr_f}
\end{equation}
with
\begin{equation}
I(m_{\tilde{\chi}_1^0}) = \int^{p_{\pi,{\rm max}}}_{140 \, \text{MeV}} \left(\frac{d\text{BR}}{dp_\pi}\right) dp_\pi \, ,
\end{equation}
where $p_{\pi,{\rm max}}$ is the maximal kinematically allowed pion
momentum for $\tilde{\chi}_1^0$s with mass $m_{\tilde{\chi}_1^0}$.  We
set $I(m_{\tilde{\chi}_1^0})$ equal to zero if
$p_{\pi,{\rm max}} < 140$ MeV.

\begin{figure*}[tbp]
  \centering
  \begin{minipage}[b]{8.1 cm}
     \includegraphics[scale=1.0]{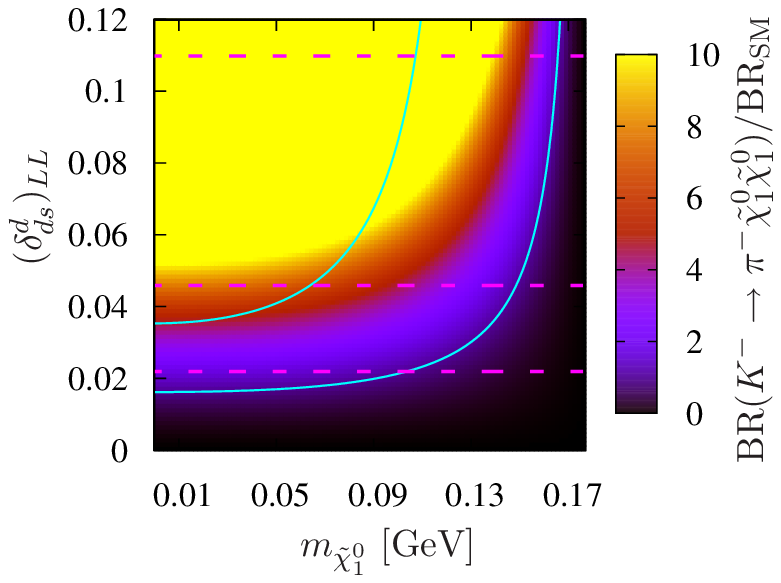}  
  \end{minipage}
  \begin{minipage}[b]{8.2 cm}
     \includegraphics[scale=1.0]{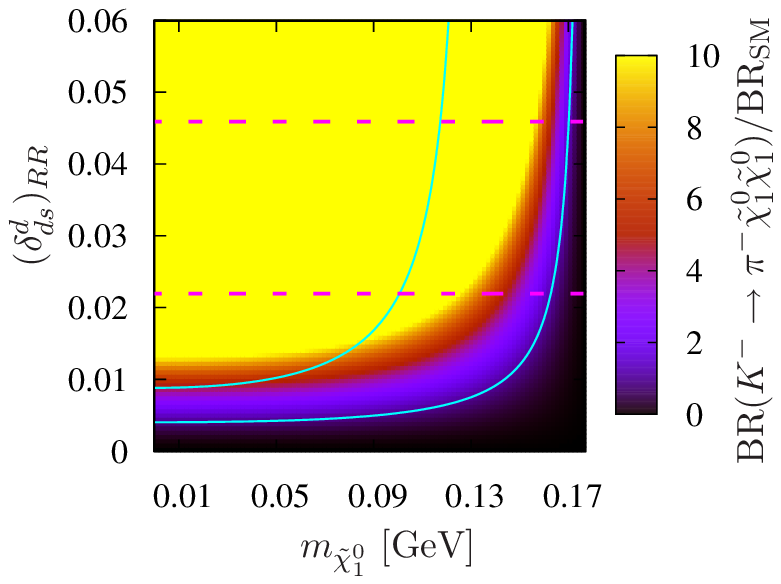}
  \end{minipage}
  \caption{\label{K_to_Pi_LL} Branching ratios (BRs) for
    $K^- \rightarrow \pi^- \tilde{\chi}_1^0 \tilde{\chi}_1^0$ as a
    function of the neutralino mass $m_{\tilde{\chi}_1^0}$ and the
    mass insertion $(\delta^d_{ds})_{LL}$ (left-hand figure) and
    $(\delta^d_{ds})_{RR}$ (right-hand figure).  The branching ratios
    are normalized to the SM prediction
    BR($K^- \rightarrow \pi^- \nu \bar \nu$)
    = $8.5 \times 10^{-11}$~\cite{SM_prediction}. We assume
    an average squark mass of $\tilde{m}=500$ GeV. The lightest gray (yellow) region
    corresponds to normalized $\text{BRs} \geq 10$. The upper (lower)
    solid gray (turquoise) line corresponds to the experimentally measured
    branching ratio plus two sigma, Eq.~(\ref{SM_exp}), multiplied
    with the correction factor $f_c$, Eq.~(\ref{corr_f}), (SM
    prediction) for $K^- \rightarrow \pi^- \nu \bar \nu$. The dashed
    lines show upper bounds on the mass insertions
    $(\delta^d_{ds})_{LL/RR}$ obtained from $K^0$--$\bar K^0$ mixing
    \cite{Ciuchini:1998ix} for different ratios of squark and gluino
    masses, {\it cf.}  Eq.~(\ref{NLO_bounds}).}
\end{figure*}

We show in Fig.~\ref{K_to_Pi_LL} the branching ratio for
$K^-\rightarrow \pi^- \tilde{\chi}_1^0 \tilde{\chi}_1^0$ as a function
of $m_{\tilde{\chi}_1^0}$ and the mass insertion parameter
$(\delta^d_{ds})_{LL}$ (left figure) and $(\delta^d_{ds})_{RR}$ (right
figure).  We assume in Eq.~(\ref{squared_ME2}) an average squark mass
of $\tilde{m}=500$ GeV. The branching ratios are normalized to the SM
prediction, Eq.~(\ref{SM_predict}).

The lower solid turquoise lines in Fig.~\ref{K_to_Pi_LL} correspond to
regions in the MSSM parameter space where the process
$K^- \rightarrow \pi^- \tilde{\chi}_1^0 \tilde{\chi}_1^0$ has a
branching ratio equal to the SM prediction, Eq.~(\ref{SM_predict}).
The upper solid turquoise lines show points in the
$m_{\tilde{\chi}_1^0}$--$(\delta^d_{ds})$ planes, where the branching
ratio for $K^- \rightarrow \pi^- \tilde{\chi}_1^0 \tilde{\chi}_1^0$ is
equal to the central experimental value for
BR($K^- \rightarrow \pi^-\nu \bar \nu$) plus $2 \sigma$,
Eq.~(\ref{SM_exp}), multiplied by the correction factor $f_c$,
Eq.~(\ref{corr_f}).  We assume, as a conservative approach, that the
branching ratio for $K^- \rightarrow \pi^- \nu \bar \nu$ is negligible
compared to the branching ratio for
$K^- \rightarrow \pi^- \tilde{\chi}_1^0 \tilde{\chi}_1^0$. With our
conservative approach, we can exclude at $2 \sigma$ the MSSM parameter
space which lies above the upper solid turquoise line. As can be seen
from Fig.~\ref{K_to_Pi_LL}, the experiments were unable to probe the
region with $m_{\tilde{\chi}_1^0} \gsim 130$ MeV, although
$\text{BR}(K^- \rightarrow \pi^- \tilde{\chi}_1^0 \tilde{\chi}_1^0)$
can be one order of magnitude larger then the respective SM process,
{\it cf.}  Eq.~(\ref{SM_predict}).  Future experiments, such as CERN
P--326/NA62 \cite{Kaon_exp}, will be able to explore roughly the MSSM
parameter space between the two solid turquoise lines as long as
$m_{\tilde{\chi}_1^0} \lsim 130$ MeV.

In Ref.~\cite{Ciuchini:1998ix}, bounds on the mass insertion
parameters $(\delta^d_{ds})_{LL/RR}$ were derived from the mass
splitting of the $K^{0}$--${\bar{K}}^{0}$ system.  For an
average squark mass of $\tilde{m}=500$ GeV, three different bounds
on $(\delta^d_{ds})_{LL/RR}$ were obtained assuming three different
ratios $x$ of gluino, $m_{\tilde{g}}$, and squark masses
\begin{eqnarray}
x \equiv m^2_{\tilde{g}}/\tilde{m}^2 = 4.0 \;\; &\Longrightarrow &\;\;
(\delta^d_{ds})_{LL/RR} \leq 1.1 \times 10^{-1} \, , \nonumber \\
x \equiv m^2_{\tilde{g}}/\tilde{m}^2 = 1.0 \;\; &\Longrightarrow &\;\;
(\delta^d_{ds})_{LL/RR} \leq 4.6 \times 10^{-2} \, , \nonumber \\
x \equiv m^2_{\tilde{g}}/\tilde{m}^2 = 0.3 \;\; &\Longrightarrow &\;\;
(\delta^d_{ds})_{LL/RR} \leq 2.2 \times 10^{-2} .
\label{NLO_bounds}
\end{eqnarray}

The horizontal dashed lines in Fig.~\ref{K_to_Pi_LL} correspond to the
bounds above.  Thus, with our calculation of the non-MFV process
$K^- \rightarrow \pi^- \tilde{\chi}_1^0 \tilde{\chi}_1^0$, we are able
for the first time to exclude the MSSM parameter space which lies
above the upper solid turquoise lines and below one of the dashed
lines.  Note that the weakest bound on $(\delta^d_{ds})_{RR}$ ($x=4$)
lies outside the right figure in Fig.~\ref{K_to_Pi_LL}.

The plot on the left in Fig.~\ref{K_to_Pi_LL} with
$(\delta^d_{ds})_{LL} \not= 0$ looks qualitatively the same as that on
the right in Fig.~\ref{K_to_Pi_LL} with $(\delta^d_{ds})_{RR} \not=
0$.  The main difference is that the branching ratio for any specific
value of $(\delta^d_{ds})$ and $m_{\tilde{\chi}_1^0}$ is $16$ times
larger in the right than in the left figure.  This is because the
light neutralino has to be bino-like and therefore couples to
hypercharge. This is larger for right-handed particles than the
respective left-handed particles in the MSSM.  We see in
Fig.~\ref{K_to_Pi_LL} that for a massless $\tilde{\chi}_1^0$, we can
strengthen the bound on the flavor-violating mass insertion parameter
$(\delta^d_{ds})_{RR}$ by a factor of between two and ten depending on
the ratio of squark mass to gluino mass, {\it cf.}
Eq.~(\ref{NLO_bounds}).

As pointed out in Sect.~\ref{nonMFV_ME}, we can also employ our
squared matrix elements, Eq.~(\ref{squared_ME2}), to calculate the
branching ratios of the flavor changing $B$ decays,
Eq.~(\ref{nonMFV_Bdecays2}). Experimentally, so far only upper bounds
for these processes exist~\cite{PDG,Aubert:2004ws,BELLE:2007zk}:
\begin{eqnarray}
\text{BR}(B^- \rightarrow \pi^- \nu \bar \nu) & \leq & 1.0 \times 10^{-4} \, , \nonumber \\ 
\text{BR}(B^- \rightarrow K^- \nu \bar \nu) & \leq & 1.4 \times 10^{-5} \, .
\label{nonMFV_Bbounds}
\end{eqnarray}  

The upper bounds on the respective mass insertion parameters are 
\begin{eqnarray}
x = 4.0 \;\;\;\; &\Longrightarrow &\;\;
(\delta^d_{db})_{LL/RR} \leq 7.0 \times 10^{-1},\;\;\;
(\delta^d_{sb})_{LL/RR} \leq 1.1 \times 10^0 \, , \nonumber \\
x = 1.0 \;\;\;\; &\Longrightarrow &\;\;
(\delta^d_{db})_{LL/RR} \leq 1.4 \times 10^{-1},\;\;\;
(\delta^d_{sb})_{LL/RR} \leq 4.8 \times 10^{-1} , \nonumber \\
x = 0.3  \;\;\;\;&\Longrightarrow &\;\;
(\delta^d_{sb})_{LL/RR} \leq 2.3 \times 10^{-1} , \nonumber \\
x = 0.25 \;\; &\Longrightarrow &\;\;
(\delta^d_{db})_{LL/RR} \leq 6.2 \times 10^{-2} .
\label{nonMFV_B_MI_bounds}
\end{eqnarray}
Again, an average squark mass of $\tilde{m}=500$ GeV is assumed.  The
bounds on $(\delta^d_{db})$ were obtained in
Ref.~\cite{Becirevic:2001jj} from the SUSY contributions to
$B_d^0$--$\bar B_d^0$ mixing and from the $B \rightarrow J/\psi \,K_s$
CP-asymmetry. We calculated the bounds on $(\delta^d_{sb})$
analogously to Ref.~\cite{Jager:2008fc} from the recently measured
$B_s^0$--$\bar B_s^0$ mass difference~\cite{PDG,Abulencia:2006ze}.

Note that additional bounds on $(\delta^d_{sb})$ from $B_s \rightarrow
X_s \ell^+ \ell^-$, $B_s \rightarrow \mu ^+ \mu^-$ and especially from
$b \rightarrow s \gamma$ exist, which can be up to an order of
magnitude greater than the bounds in Eq.~(\ref{nonMFV_B_MI_bounds}).
In addition, SUSY contributions to $B_s^0$--$\bar B_s^0$ mixing can be
dominated by two-loop double Higgs penguins for large $\tan \beta$.
For more details, see for example
Refs~\cite{Jager:2008fc,Foster:2006ze,Ciuchini:2002uv,
Silvestrini:2007yf,Foster:2005kb} and references therein. However,
these bounds are highly model dependent, {\it i.e.} they depend not
only on the squark and gluino masses. We will therefore not consider
them in our analysis.

\begin{figure*}[tbp]
  \centering
  \begin{minipage}[b]{8.1 cm}
      \includegraphics[scale=1.0]{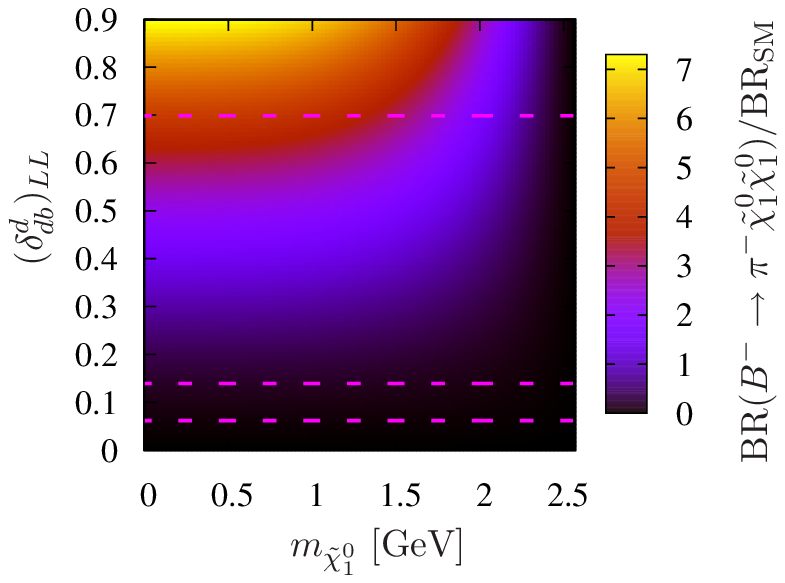}  
  \end{minipage}
  \begin{minipage}[b]{8.2 cm}
      \includegraphics[scale=1.0]{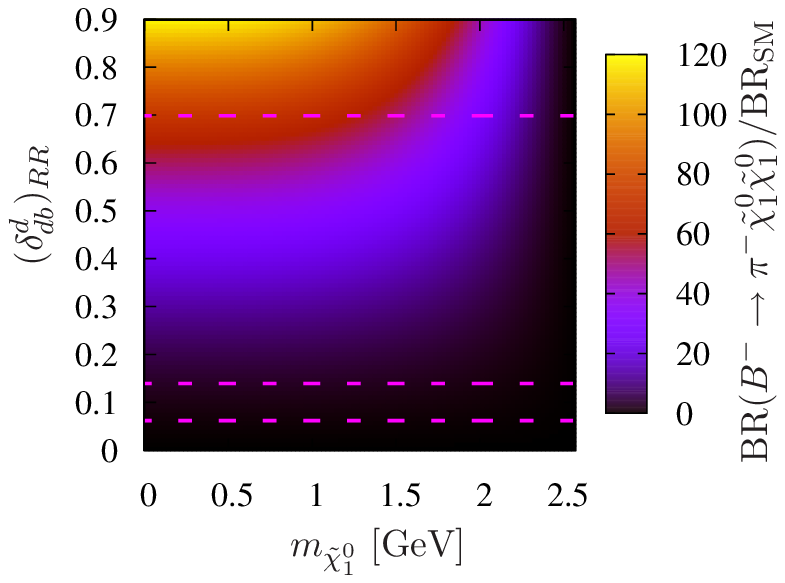}
  \end{minipage}
  \caption{\label{B_to_Pi} Branching ratios for
    $B^- \rightarrow \pi^- \tilde{\chi}_1^0 \tilde{\chi}_1^0$ as a
    function of the neutralino mass $m_{\tilde{\chi}_1^0}$ and the mass
    insertion $(\delta^d_{db})_{LL}$ (left-hand side) and
    $(\delta^d_{db})_{RR}$ (right-hand side).  The branching ratio is
    normalized to the calculated BR for $B^{-} {\to} {\pi}^{-} {\nu} {\bar{{\nu}}}$ within the SM, which is $2.2 {\times} 10^{-7}$ \cite{Jeon:2006nq}. We assume an average squark mass of
    $\tilde{m}=500$ GeV. The dashed lines show upper bounds on the
    mass insertions $(\delta^d_{db})_{LL/RR}$ mainly obtained from
    $B_d^0$--$\bar B_d^0$ mixing for different ratios of squark mass
    to gluino mass \cite{Becirevic:2001jj},
    {\it cf.} Eq.~(\ref{nonMFV_B_MI_bounds}).}
\end{figure*}    

We show in Fig.~\ref{B_to_Pi} the branching ratios for
$B^- \rightarrow \pi^-\tilde{\chi}_1^0 \tilde{\chi}_1^0$ as a function
of $m_{\tilde{\chi}_1^0}$ and the mass insertions $(\delta^d_{db})_{LL}$
(left figure) and $(\delta^d_{db})_{RR}$ (right figure). The dashed
lines correspond to the upper bounds on the the mass insertions as
given in Eq.~(\ref{nonMFV_B_MI_bounds}). The allowed region for
non-vanishing $(\delta^d_{db})_{LL}$ [$(\delta^d_{db})_{RR}$],
{\it i.e.} the region below the dashed lines, lies at least two [one]
order of magnitude below the current experimental upper bound of $1.0
\times 10^{-4}$, {\it cf.} Eq.~(\ref{nonMFV_Bbounds}). We can thus not
exclude additional regions of the MSSM parameter space from the upper
bound on the branching ratio. However, a future super $B$ factory will
be able to explore large parts of the parameter space shown in
Fig.~\ref{B_to_Pi} \cite{superB}, {\it i.e.} branching ratios down to
$\mathcal{O}(10^{-6}-10^{-7})$.  Note that if
${\mxone} \gsim 1$~GeV, the momentum of the pion or kaon from the
$B$ meson decay may be too soft to pass experimental cuts.

\begin{figure*}[tbp]
  \centering
  \begin{minipage}[b]{8.1 cm}
      \includegraphics[scale=1.0]{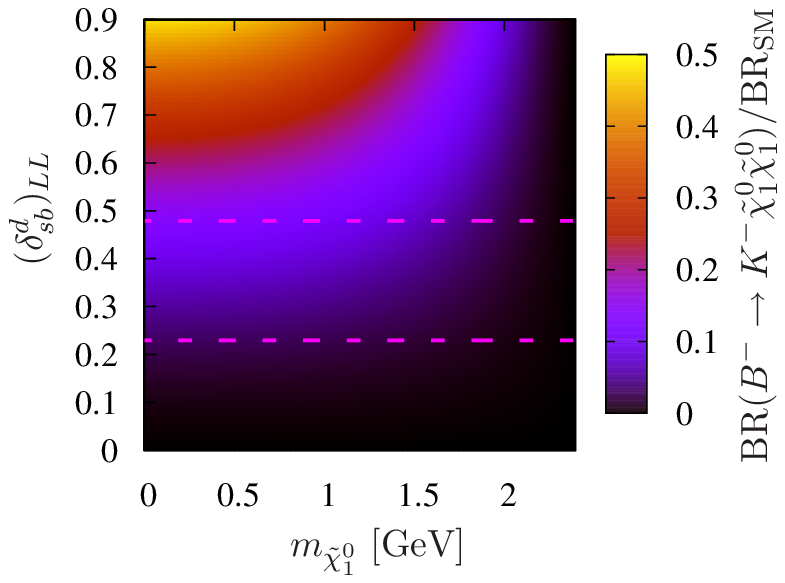}  
  \end{minipage}
  \begin{minipage}[b]{8.2 cm}
      \includegraphics[scale=1.0]{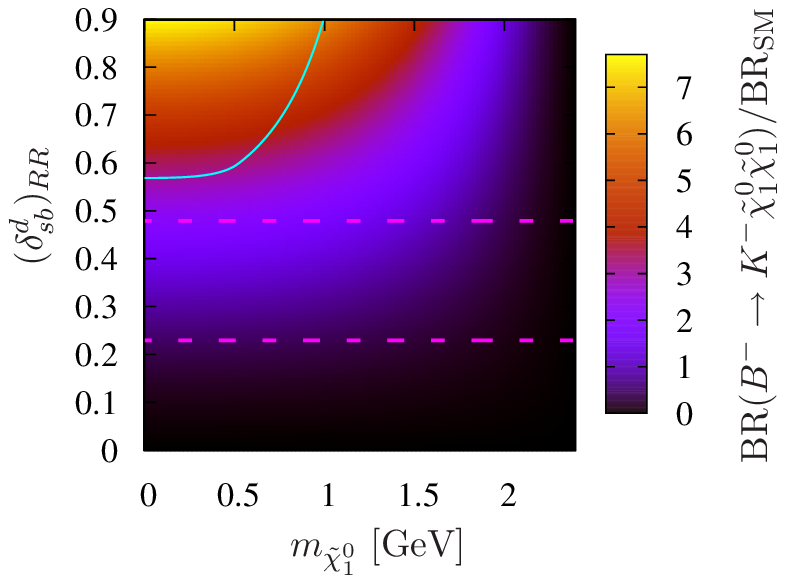}
  \end{minipage}
  \caption{\label{B_to_K} Branching ratios for
    $B^- \rightarrow K^- \tilde{\chi}_1^0 \tilde{\chi}_1^0$ as a
    function of the neutralino mass $m_{\tilde{\chi}_1^0}$ and the
    mass insertion $(\delta^d_{sb})_{LL}$ (left-hand side) and
    $(\delta^d_{sb})_{RR}$ (right-hand side).  The branching ratio
    is normalized to the calculated BR for $B^{-} {\to} K^{-} {\nu} {\bar{{\nu}}}$ within the SM, which is $4.5 {\times} 10^{-6}$ \cite{Buchalla:2000sk, Altmannshofer:2009ma}. We assume an average squark mass
    of $\tilde{m} = 500$ GeV. The solid gray (turquoise) line
    corresponds to the experimental upper bound,
    Eq.~(\ref{nonMFV_Bbounds}), multiplied by a correction factor
    analogously to Eq.~(\ref{corr_f}). The dashed lines show
    upper bounds on the mass insertions $(\delta^d_{sb})_{LL/RR}$
    obtained from $B_s^0$--$\bar B_s^0$ mixing for different
    ratios of squark mass to gluino mass,
    {\it cf.}~Eq.~(\ref{nonMFV_B_MI_bounds}).  The weakest bound
    on $(\delta^d_{sb})$ ($x=4$) lies outside the figure.}
\end{figure*}  

Finally we show in Fig.~\ref{B_to_K} the branching ratios for the
process $B^- \rightarrow K^- \tilde{\chi}_1^0 \tilde{\chi}_1^0$ as a
function of $m_{\tilde{\chi}_1^0}$ and $(\delta^d_{sb})_{LL/RR}$. The
numerical values are of the same order of magnitude as those in
Fig.~\ref{B_to_Pi}. Differences are mainly due to different form
factors, see Eq.~(\ref{nonMFV_Bform_factors}). In contrast to the $B$
decay into a pion, the decay into a kaon has a stricter upper bound on
the respective branching ratio, namely $1.4 \times 10^{-5}$, {\it cf.}
Eq.~(\ref{nonMFV_Bbounds}). The solid turquoise line in
Fig.~\ref{B_to_K} corresponds to this upper bound multiplied by a
correction factor analogous to Eq.~(\ref{corr_f}). We have taken into
account that the BELLE collaboration required the kaon momentum to lie
within a window of $1.6$ GeV and $2.5$ GeV \cite{BELLE:2007zk}. We can
therefore exclude the MSSM parameter space above the turquoise line.
Again, with a future super $B$ factory we will be able to explore
large regions of the parameter space in Fig.~\ref{B_to_K}
\cite{superB}.

To conclude this section, let us briefly comment on the validity of
the mass insertion approximation, Eq.~(\ref{mass_expansion}).  For the
decay $K^- \rightarrow \pi^- \tilde{\chi}_1^0 \tilde{\chi}_1^0$, the
relevant mass insertion parameters are of the order
$\delta^d_{ds}=\mathcal{O}(10^{-2})$ and an expansion in
$\delta^d_{ds}$ converges very fast. For the $B$ decays, however, we
have considered mass insertions of up to $\delta^d_{db/sb}= 0.9$,
{\it cf.}  Figs.~\ref{B_to_Pi} and \ref{B_to_K}. While it is clear
that the analysis should eventually be refined by taking into account
higher order terms in the expansion Eq.~(\ref{mass_expansion}), our
approximation is sufficient to demonstrate that $B$ decays into very
light $\tilde{\chi}_1^0$ can be tested with the next generation of
$B$ factories.

\section{Light neutralino production at hadron colliders}
\label{sec:monojet}

A massless or very light ${\xone}$ that is purely bino is not
fundamentally different in terms of collider signals than scenarios
where it is ${\mathcal{O}}( 100$~GeV) and mainly bino, such as SPS1a
($97$ GeV, $98.5\%$ bino).  The main effect is that the phase space is
bigger, in general increasing cross-sections, especially where the
energy is not much greater than the creation threshold.  There are
many signals of interest at colliders, not least those involving decay
chains, but the masslessness of the lightest neutralino does not
fundamentally change those analyses.

It is reasonable to ask whether the increase due to the greater phase
space is in conflict with searches at colliders so far.  The bino component does not couple to the $Z$ boson at tree level and so constraints from the invisible width of the $Z$ are only applicable to the very small higgsino components and also to the bino component at one-loop level.  No
bounds on the neutralino mass can thus be deduced from the $Z$
width~\cite{Dreiner:2007fw,Dreiner:2009ic,Choudhury:1999tn}.  Direct
searches at past and future $e^{+} e^{-}$ colliders have been studied
in detail in Refs.~\cite{Fayet:1982ky, Ellis:1982zz, Choudhury:1999tn,
Dedes:2001zia,Dreiner:2006sb, Dreiner:2007vm}.  While no bounds on the
mass of the lightest neutralino can be obtained from past searches at
LEP and $B$ factories, measurements at a future $e^{+} e^{-}$ linear
collider might be able to discover light neutralinos through radiative
production, $e^{+} e^{-} {\to} {\xone} {\xone} + {\gamma}$.

Direct hadroproduction of light neutralinos in association with a jet,
\begin{equation}
  p {\bar{p}} / p p {\to} {\xone} {\xone} + {\rm jet},
\label{eq:monojet}
\end{equation}
will lead to a spectacular monojet signature. We have calculated the
tree-level cross-section for the pair production of massless $\xone$
plus one jet, Eq.~(\ref{eq:monojet}), for both the Tevatron
(${\sqrt{s}} = 1.96$ TeV) and the LHC (${\sqrt{s}} = 14$ TeV) using
\MadGraph~\cite{MadGraph}.  We require a jet with transverse
momentum of at least $80$~GeV.  The numerical results are collected in
Table~\ref{monojet_table} for the pseudo-SPS points described in
Sect.~\ref{MFV_section} and are compared to the original SPS points
with ${\xone}$ masses of typically ${\mathcal{O}} (100$~GeV).  Even
though the Tevatron cross sections for massless ${\xone}$ are enhanced
by a factor up to $2.5$ with respect to the original SPS
cross-sections with massive neutralinos, the expected number of events
is only about $2$ or less for the full $6$ fb${}^{-1}$ of integrated
luminosity of Run II so far.  Given the large SM backgrounds, such as
$p {\bar{p}} {\to} Z ( {\to} {\nu} {\bar{{\nu}}} ) + {\rm jet}$, we
conclude that current and future monojet searches at the
Tevatron~\cite{Abazov:2003gp, Aaltonen:2008hh} will not be sensitive
to the direct pair-production of light neutralinos.  Similar
conclusions hold for radiative production with an additional photon,
$p {\bar{p}} {\to} {\xone} {\xone} + \gamma$.

At the LHC, light neutralino pair production with jets is only
enhanced from the massive case by about $20\%$.  Still, with
cross-sections of ${\mathcal{O}} (100$~fb), see
Table~\ref{monojet_table}, detection of these processes should be
possible with sufficiently high luminosity and an excellent
understanding of SM backgrounds~\cite{Vacavant:2001sd,Weng:2006bc}.

\begin{center}
\begin{table}[ht]
\begin{tabular}{l d@{${\columnspacefixer}$}l@{\hspace{1cm}} 
d@{${\columnspacefixer}$}l@{\hspace{1cm}} 
d@{${\columnspacefixer}$}l@{\hspace{1cm}} 
d@{${\columnspacefixer}$}l}
\hline \\[-9mm] \hline
(pseudo-) & \multicolumn{4}{c}{Tevatron cross-section} & \multicolumn{4}{c}
{LHC cross-section}\\
SPS & \multicolumn{2}{c}{pseudo-SPS} & \multicolumn{2}{c}{normal SPS} & 
\multicolumn{2}{c}{pseudo-SPS} & \multicolumn{2}{c}{normal SPS}
\newhline
1a  & $2.96$ & ${\times} 10^{-1}$ fb & $1.68$ & ${\times} 10^{-1}$ fb & $3.26$ & ${\times} 10^{+2}$ fb & $2.73$ & ${\times} 10^{+2}$ fb
\\
2   & $2.88$ & ${\times} 10^{-3}$ fb & $1.99$ & ${\times} 10^{-3}$ fb & $3.25$ & ${\times} 10^{+2}$ fb & $3.01$ & ${\times} 10^{+2}$ fb
\\
3   & $2.66$ & ${\times} 10^{-2}$ fb & $7.75$ & ${\times} 10^{-3}$ fb & $5.52$ & ${\times} 10^{+1}$ fb & $4.61$ & ${\times} 10^{+1}$ fb
\\
4   & $4.48$ & ${\times} 10^{-2}$ fb & $1.84$ & ${\times} 10^{-2}$ fb & $8.59$ & ${\times} 10^{+1}$ fb & $7.14$ & ${\times} 10^{+1}$ fb
\\
5   & $9.18$ & ${\times} 10^{-2}$ fb & $3.78$ & ${\times} 10^{-2}$ fb & $1.54$ & ${\times} 10^{+2}$ fb & $1.32$ & ${\times} 10^{+2}$ fb
\newhline \\[-9mm] \hline
\end{tabular}
\normalsize
\caption{Monojet cross-sections ${\sigma}( p {\bar{p}} / p p {\to} {\xone} {\xone} + {\rm jet} )$ at the Tevatron and at the LHC.  Shown are results for the pseudo-SPS points with massless ${\xone}$ in comparison to the original SPS cross sections with massive neutralinos.}
\label{monojet_table}
\end{table}
\end{center}

\section{Conclusion}
\label{sec:conclusion}

Rare meson decays provide a sensitive test for the presence of new
physics. In this paper we have studied the decay of pseudoscalar and
vector mesons to light neutralinos in the MSSM.  We have presented
details of the calculations and explicit formulae for the two-body
decays $M {\to} {\xone} {\xone}$, where
$M = {\pi}^{0}, {\eta}, {{\eta}'}, {\rho}^{0}, {\omega}, {\phi},
J/{\psi}, {\Upsilon}$.  Furthermore, we have
performed the first complete calculation of the loop-induced decays
${\Ktopi}$ and ${\BtoK}$. Considering various MSSM scenarios we find
that the supersymmetric branching ratios are several orders of
magnitude smaller then the SM processes with neutrinos instead of
neutralinos in the final state.  Consequently, no bounds on the
neutralino mass can be inferred from rare meson decays in the MSSM
with minimal flavor violation.  However, the branching ratios for the
${\Ktopi}$ and ${\BtoK}$ decays may be significantly enhanced when one
allows for non-minimal flavor violation.  We find new constraints on
the MSSM parameter space for such scenarios and argue that future
experiments in the kaon and $B$ meson sector may be able probe light
neutralinos with masses up to approximately $1$~GeV from rare
decays.  We have also considered searches for light neutralinos from
monojet signatures at hadron colliders.  While current and future
monojet searches at the Tevatron will not be sensitive to the direct
pair-production of light neutralinos, the detection of these processes
should be possible at the LHC with sufficiently high luminosity and an
excellent understanding of SM backgrounds.

\begin{acknowledgments}
  We want to thank JoAnne Hewett, Gudrun Hiller, Olaf Kittel,
  Christopher Smith and Bryan Webber for discussions and suggestions.
  This work is supported in part by the European Community's
  Marie-Curie Research Training Network under contract
  MRTN-CT-2006-035505 ``Tools and Precision Calculations for Physics
  Discoveries at Colliders'', the DFG SFB/TR9 ``Computational Particle
  Physics'' and the Helmholtz Alliance ``Physics at the Terascale''.
  S.G.~thanks the Deutsche Telekom Stiftung and the Bonn-Cologne
  Graduate School of Physics and Astronomy for financial support.
  D.K.~thanks the Queen Mary, University of London Gradute school for
  financial support.
\end{acknowledgments}

\appendix

\section{Form Factors for
$K^- \rightarrow \pi^- \tilde{\chi}_1^0 \tilde{\chi}_1^0$
and
$B^- \rightarrow~K^-/\pi^- \tilde{\chi}_1^0 \tilde{\chi}_1^0$}
\label{form_factors}

For the decay
$K^- \rightarrow \pi^- \tilde{\chi}_1^0 \tilde{\chi}_1^0$,
the quark currents need to be replaced by their
corresponding hadronic matrix elements involving a kaon $K^-$ and a
pion $\pi^-$\cite{PDG}:
\begin{eqnarray}
\left< \pi^-(p_{\pi})|\bar d \gamma^\mu P_{L/R} s|K^-(p_K) \right> 
 = {\frac{1}{{2}}} \left[ f_+(t) (p_K + p_\pi )^\mu + f_-(t) (p_K - p_\pi )^\mu \right] \, ,
\label{Kl3_form_factor}
\end{eqnarray}   
and
\begin{eqnarray}
f_-(t) = [f_0(t) - f_+(t)] \frac{(m_k^2-m_\pi^2)}{t} \, ,
\end{eqnarray}
where $t$ is defined via $t=(p_K-p_\pi)^2$.  The functions
$f_+$ and $f_0$ are the so-called $K_{l3}$ form factors.  We describe them
with the help of the linear approximation \cite{PDG}
\begin{eqnarray}
f_{+/0}(t)=f_{+/0}(0) \left( 1 + \lambda_{+/0} \frac{t}{m^2_\pi} \right) \, ,
\end{eqnarray}  
with $\lambda_+=2.96 \times 10^{-2}$, $\lambda_0=1.96 \times 10^{-2}$ and
$f_{+}(0) = f_{0}(0) = 1.013$, where we calculated $f_{+}(0)$ using the
measured branching ratio for $K^- \rightarrow \pi^0 e^- \bar\nu_e$ and
isospin symmetry, {\it i.e.}
\begin{eqnarray}
 \left< \pi^-(p_{\pi})|\bar d \gamma^\mu P_{L/R} s|K^-(p_K) \right>
= \sqrt{2} \left< \pi^0(p_{\pi})|\bar u \gamma^\mu P_{L/R} s|K^-(p_K) \right> \, .
\end{eqnarray}

The decays
\begin{equation}
B^- \rightarrow \pi^-/ K^- \tilde{\chi}_1^0 \tilde{\chi}_1^0 \, , 
\label{nonMFV_Bdecays}
\end{equation} 
can be described in a similar way. One has to replace in
Eq.~(\ref{Kl3_form_factor}) $p_K$ by $p_B$ and, if one is
considering ${\BtoK}$, $p_\pi$ by $p_K$.  For the parametrization
of the respective form factors $f^{\pi/K}_+$ and $f^{\pi/K}_0$, we
use \cite{Ball:2004hn,Ball:2004ye}
\begin{eqnarray}
f^{\pi}_+ &=& \frac{r_1^{\pi}}{1-t/(m_1^{\pi})^2} + \frac{r_2^{\pi}}{1-t/
(m_2^{\pi})^2}\, , \nonumber \\
f^{K}_+ &=& \frac{r_1^{K}}{1-t/(m_1^{K})^2} + \frac{r_2^{K}}{[1-t/(m_1^{K})
^2]^2}\, , \nonumber \\
f^{\pi/K}_0 &=& \frac{r_3^{\pi/K}}{1-t/(m_3^{\pi/K})^2}\, , 
\label{nonMFV_Bform_factors}
\end{eqnarray} 
where $m_1^\pi=5.32$ GeV and $m_1^K=5.41$ GeV are the masses of the
$B^*$ and $B_s^*$ vector mesons, respectively.  The fit parameters are
$r_1^\pi=0.744$, $r_2^\pi=-0.486$, $(m_2^\pi)^2=40.73$ $\text{GeV}^2$,
$r_1^K=0.162$, $r_2^K=0.173$, $r_3^\pi=0.258$, $r_3^K=0.330$,
$(m_3^\pi)^2=33.81$ $\text{GeV}^2$ and $(m_3^K)^2=37.46$
$\text{GeV}^2$.

Using this parameterization for the non-MFV ${\Ktopi}$ decay,
we obtain for the squared matrix elements summed over final state spins
\begin{equation}
{\overline{{|{\mathcal{M}}_{LL/RR}|^{2}}}} =
C_{LL/RR} \frac{e^4 (\delta^d_{ds})^2_{LL/RR}}{\cos^4 \theta_W \tilde{m}^4}
\left( {\overline{{{\mathcal{M}}^{2}_{++}}}} + 
{\overline{{{\mathcal{M}}^{2}_{+-}}}} +
{\overline{{{\mathcal{M}}^{2}_{--}}}} \right) \, ,
\label{squared_ME2}
\end{equation}
with
\begin{eqnarray}
{\overline{{{\mathcal{M}}^{2}_{++}}}} &=&
4 f_+^2 [ m^2_{\tilde{\chi}_1^0}(m^2_{\tilde{\chi}_1^0} + m^2_\pi) 
+ (k_1 \cdot k_2) (m^2_{\tilde{\chi}_1^0} - m_\pi^2) \nonumber \\
& & + 2 m^2_{\tilde{\chi}_1^0} (k_1 \cdot p_\pi) + 2 m^2_{\tilde{\chi}_1^0} 
(k_2 \cdot p_\pi)\nonumber \\
& &+ 2 (k_1 \cdot p_\pi) (k_2 \cdot p_\pi) ] \, , \nonumber \\
{\overline{{{\mathcal{M}}^{2}_{+-}}}} &=& 8 f_+ f_- m^2_{\tilde{\chi}_1^0} 
[ m^2_{\tilde{\chi}_1^0} 
+ (k_1 \cdot k_2) + (k_1\cdot p_\pi) + (k_2\cdot p_\pi)] \, , \nonumber \\
{\overline{{{\mathcal{M}}^{2}_{--}}}} &=& 4 f_-^2 [ m^2_{\tilde{\chi}_1^0} 
(k_1 \cdot k_2) + m^4_{\tilde{\chi}_1^0} ] \, ,
\label{squared_ME}
\end{eqnarray}
and the constants $C_{LL} = 1 / 1296$ and $C_{RR} = 1 / 81$.  The
large difference between $C_{LL}$ and $C_{RR}$ originates from the
different hypercharges of right- and left-handed squarks, with
which they couple to the bino-like $\tilde{\chi}_1^0$.


\end{document}